\documentclass[twocolumn]{autart}    

\usepackage{graphicx}          
                               
\usepackage[mathscr]{eucal}
\usepackage{array}
\usepackage{epsfig}
\usepackage{amsfonts}
\usepackage{amsmath}
\usepackage{amssymb}
\usepackage{amsxtra}
\usepackage{varioref}
\usepackage{color}
\usepackage{enumerate}
\usepackage{epstopdf}
\usepackage{mathtools} 
\usepackage{bm}

\newcommand{\fin}{\hfill{$\square$}}
\newcommand{\qedd}{\hfill{$\blacksquare$}}
\newcommand{\SINR}{\text{SINR}}
\newtheorem{lemma}{Lemma}
\newtheorem{remark}{Remark}
\newtheorem{proposition}{Proposition}
\newtheorem{assu}{Standing Assumption}
\newtheorem{theorem}{Theorem}

\newtheorem{corollary}{Corollary}
\newtheorem{problem}{Problem}

\newcommand{\w}{\bm{\lambda}}
\newcommand{\mtxdu}[2]{\begin{bmatrix} #1\\ #2 \end{bmatrix}}

\newcommand{\mtxdd}[4]{\begin{bmatrix} #1 & #2\\ #3 & #4 \end{bmatrix}}

\newcommand{\smallmtx}{\setlength\arraycolsep{3pt}
	\def\arraystretch{1}}
\newcommand{\non}{\nonumber}
\newcommand{\R}{\mathbb{R}}
\newcommand{\N}{\mathbb{N}}

\newcommand{\diag}{\mbox{diag}}
\newcommand{\esp}[1]{\mathbb{E}\{#1\}}
\newcommand{\espp}[1]{\mathbb{E}\left\{#1\right\}}
\newcommand{\norm}[1]{\|#1\|}
\newcommand{\EL}{\mathcal{L}}

\newcommand{\LTI}{\text{\tiny LTI}}

\begin{document}
\begin{frontmatter}

\title{Transmit power policies for stochastic stabilisation of multi-link wireless networked control systems\thanksref{footnoteinfo}} 

\thanks[footnoteinfo]{This work was funded by the Australian Research Council under the Discovery Grant DP200101303. Corresponding author A.I.~Maass.}

\author[unimelb]{Alejandro I. Maass}\ead{maassa@unimelb.edu.au},    
\author[unimelb]{Dragan Nesic}\ead{dnesic@unimelb.edu.au},
\author[cran]{Romain~Postoyan}\ead{romain.postoyan@univ-lorraine.fr},               
\author[cran]{Vineeth~S.~Varma}\ead{vineeth.satheeskumar-varma@univ-lorraine.fr},  
\author[cran]{Samson~Lasaulce}\ead{samson.lasaulce@univ-lorraine.fr}

\address[unimelb]{Department of Electrical and Electronic Engineering, The University of Melbourne, Parkville, 3010, Victoria, Australia}  
\address[cran]{Universit\'e de Lorraine, CNRS, CRAN, F-54000 Nancy, France}


\begin{abstract}                          
Transmit power control is one of the most important issues in wireless networks, where nodes typically operate on limited battery power. Reducing communicating power consumption is essential for both economic and ecologic reasons. In fact, 
transmitting at unnecessarily high power not only reduces node lifetime, but also introduces excessive interference and electromagnetic pollution. Existing work in the wireless community mostly focus on designing transmit power policies by taking into account communication aspects like quality of service or network capacity. Wireless networked control systems (WNCSs), on the other hand, have different and specific needs such as stability, which require transmit power policies adapted to the control context. Transmit power design in the control community has recently attracted much attention, and available works mostly consider linear systems or specific classes of non-linear systems with a single-link view of the system.
In this paper, we propose a framework for the design of stabilising transmit power levels that applies to much larger classes of non-linear plants, controllers, and multi-link setting. By exploiting the fact that channel success probabilities are related to transmit power in a non-linear fashion, we first derive closed-loop stability conditions that relate channel probabilities with transmission rate. Next, we combine these results together with well-known and realistic interference models to provide a design methodology for stabilising transmit power in non-linear and multi-link WNCSs.
\end{abstract}
\end{frontmatter}

	\section{Introduction}
	Wireless networked control systems (WNCSs) are composed of spatially distributed sensors, actuators, and controllers communicating via wireless networks rather than point-to-point wired connections. WNCSs are increasingly used in a range of Industrial Internet of Things (IIoT) application domains such as vehicle-to-vehicle communications in automated highways \cite{dey2016vehicle}, and complex industrial settings 	\cite{ahaker19}. 
	In WNCSs, where nodes are likely to operate on limited battery life, energy saving is a crucial issue \cite{rault2014energy}. Indeed, the wireless communication cost at many operational modes, such as transmit and receive, is a dominating factor affecting battery power consumption. To prolong the lifetime of a node and avoid unnecessary maintenance, wireless transmit power control becomes essential for WNCSs \cite{pantazis2007survey}.
	In fact, transmitting at excessively high power is inefficient because of mutual interference between nodes and the limited battery lifetime. Ideally, transmit power of a node should be adjusted on a link-by-link basis to improve energy efficiency \cite{elbatt2004joint}. We highlight that, in the  wireless  literature,  various  studies have investigated transmit power control as a solution to interference management for communication purposes such as optimising quality of service (QoS), maximizing transmission rate, packet success rate, or energy-efficiency, see e.g. \cite{rault2014energy}  for  an  extensive  survey. While these works are fully relevant in the context of wireless networks, they focus on communication objectives only, and thus are not---\emph{a priori}---well suited for WNCSs, where different control-oriented requirements are needed, e.g. closed-loop stability and state estimation.
	
	The study of transmit power in the context of WNCSs has recently attracted much attention in the control community, see e.g. \cite{quahle12,rewujo17,liwuch18,pezzutto2021transmission} for state estimation results, and \cite{garipa14,vaolpo19,varma2020time,lima2020resource,balaghiinaloo2020lq} for stabilisation, where the latter are more relevant to the work herein. Particularly, \cite{garipa14} deals with stability and power control for linear plants over a single-link wireless network. The work \cite{vaolpo19} considers the control of linear plants by an output feedback law over a single-link wireless network, where the objectives are both mean square stabilisation and minimising the expected average communication energy. 
	Also for linear and single-link WNCSs, \cite{balaghiinaloo2020lq} considers performance by jointly designing (state dependant) transmit power and controller in order to minimise an average quadratic cost.
	Later on, \cite{varma2020time} provided policies for stability under minimum average power usage for single-link non-linear WNCSs. Lastly, \cite{lima2020resource}, provides a reinforcement learning framework for both stability and power control also for single-link non-linear WNCSs. 
	To better improve energy efficiency in WNCSs, transmit power control strategies should take into account the multiple wireless links that compose a WNCS, as opposed to using a single-link view. Currently, there are no available results on transmit power policies for stabilisation of general non-linear plants over multiple wireless links.

	To derive stabilising power control policies in such setting, we will exploit the fact that probabilities of success in each wireless link depend, in a non-linear fashion, upon the transmit power levels. It is then essential to first derive stability conditions---in terms of channel probabilities of success---for the considered non-linear multi-link setting, and then build upon these to design transmit power levels. We note that available stability works on WNCSs subject to packet losses are not directly applicable to the considered setting, and thus cannot be used off-the-shelf for our purposes. Indeed, initial works, such as \cite{scsifr07,mofoso13,moobas15,lisush17}, have studied the single-link case for linear systems in which only one probability determines the success of the transmission. For single-link non-linear systems, a predictive control strategy is developed in \cite{quenes12} to stabilise discrete-time plants under Markovian packet dropouts.
	An extra modelling layer where multiple links are in place has been considered for non-linear systems in \cite{hestee06,tabnes08-tac,antunes2013stability}.
	Even though the latter cover multiple links, it is assumed that a single probability determines the transmission success of all these links, which is often unrealistic.
	Multiple links with different probabilities of successful reception have been thus far considered only for linear systems in \cite{gasigo12,cheqiu15,ricel15,mavasi16}, and to the best of our knowledge, there are no stability results for general non-linear systems in this context. 
	%

	
	
	In response to the above discussion, this paper presents a framework to design transmit power levels for general non-linear plants where
	communications are done via a wireless network with multiple links subject to random packet loss with different probabilities of success for each wireless link. 
	As mentioned above, the first step is to derive stability results in such setting, which will later be exploited to obtain stabilising transmit power policies. Our methodology is based on the emulation approach \cite{wabebu01,nestee04}, i.e. we first construct a stabilising controller while ignoring the wireless network, and we next take the network into account to establish sufficient conditions that preserve the original stability property. Specifically, our main contributions are the following.
	\begin{itemize}
		\item We first provide a stochastic hybrid model for the WNCS that captures both the continuous dynamics from the plant/controller, and the discrete dynamics from the wireless links such as random packet losses and random transmission instants. This model is significantly more general than previous discrete-time linear models considered in the literature, e.g. \cite{gasigo12,cheqiu15,ricel15,mavasi16}. With respect to the non-linear works \cite{tabnes08-tac,hestee06}, we consider a more general case that allows for different probabilities of success for each wireless link, which leads to new analysis. 
		\item For this stochastic hybrid model, we provide sufficient conditions on the transmission rate that ensure stability of the WNCS in this broader multi-channel setting. These results illustrate how channel probabilities affect stability. The main stability property we use throughout the paper is $\EL_p$ stability in expectation, and we also provide additional stability properties for disturbance-free plants such as exponential stability in expectation and in probability. 
		The stability results are provided for two classes of scheduling protocols, i.e. \emph{deterministic protocols} (e.g., round robin \cite{ugrinovskii2014round} and TDMA-like protocols \cite{cionca2008tdma}) and \emph{stochastic protocols} (e.g., CSMA-like protocols \cite{chen2010wirelesshart}), and also for linear time-invariant systems (LTI) as a special case. The LTI stability results are given in terms of LMI-type conditions. 
		\item We show that, when applicable, previous stability results like  \cite{tabnes08-tac} are demonstrably more conservative since they do not respect the multi-link nature of the problem. The obtained stability results serve as an extension of previous works on WNCSs subject to packet losses that were presented either for the single-link case or multi-link case with a single probability of success for every channel.
		\item Lastly, and importantly, we exploit the obtained stability results to provide a methodology to determine transmit power levels that achieve stability of the WNCS.
		Moreover, by using the stability condition as a constraint, we propose an optimisation problem that minimises total power whilst ensuring stability of the WNCS. Different costs can be potentially used depending on the application. We show that for the case of designing transmit power levels for one cluster with multiple links, the considered optimisation problem can be solved via multiple linear programs (LPs) that are computationally inexpensive. Interestingly, LPs have already appeared in the wireless community for power control problems that do not take stability into account, instead, these are interested in optimising QoS \cite{chiang2008power}. Therefore, we believe our results help bridging the gap between control and communications by designing power control policies that not only take stability of the WNCS into account, but also use well-known and realistic interference models. 
	\end{itemize}


\emph{Notation:}
Let $\N\coloneqq  \{0,1,2,\dots\}$, $\N_{>0}\coloneqq  \N\setminus\{0\}$, $\R^{m\times n}$ the set of all real matrices of dimension $m\times n$, $n,m\in\N_{>0}$, $\R_{\geq 0}\coloneqq  [0,\infty)$, and $\R_{>0}\coloneqq  (0,\infty)$.
For any $x\in\R^n$ and $y\in\R^m$, we use $(x,y)\coloneqq  [x^\top \ y^\top]^\top\in\R^{n+m}$.
For $x\in\R^n$, $|x|$ denotes the standard Euclidean norm, and also the induced 2-norm for a real matrix. For $x=(x_1,\dots,x_n)\in\R^n$, $\overline{x}=(|x_1|,\dots,|x_n|)$. For any matrix $M$, the entries of $\overline{M}$ are the absolute values of the corresponding entries of $M$. The identity matrix of dimension $n\times n$ is denoted by $I_n$. $\diag\{M_1,\dots,M_N\}$ returns the block diagonal matrix with the matrices $M_1,\dots,M_N$ along the diagonal. Let $x=(x_1,\dots,x_n)\in\R^n$ and $y=(y_1,\dots,y_n)\in\R^n$, a partial order $\preceq$ is given by $x\preceq y\Leftrightarrow x_i\leq y_i$, for all $i\in\{1,\dots,n\}$. The expectation and probability operators are denoted by $\esp{\cdot}$ and $\Pr\{\cdot\}$, respectively. 
Given a (Lebesgue) measurable function $f:\R \rightarrow \R^n$, $\norm{f}_{\EL_p}\coloneqq  \left(\int_{\R} |f(s)|^p\,\mathrm{d}s \right)^{1/p}$,
for $p\in\N_{>0}$, $\norm{f}_{\EL_{\infty}}\coloneqq  \text{ess\,sup}_{t\in\R}\, |f(t)|$, and $\norm{f}_{\EL_{\infty}[a,b]}\coloneqq  \text{ess\,sup}_{t\in[a,b]}\, |f(t)|$. 
We write $f\in\EL_p$ for $p\in\N_{>0}\cup\{+\infty\}$ whenever $\norm{f}_{\EL_p}<\infty$. 
Given $[a,b]\subset \R$, $\norm{f}_{\EL_p[a,b]}\coloneqq  \big(\int_{[a,b]} |f(s)|^p\,\mathrm{d}s \big)^{1/p}$ denotes the $\EL_p$ norm of $f$ when restricted to the interval $[a,b]$. For a measurable function $g:\Omega\times \R  \rightarrow \R^n$, we say that $g\in\EL_p^e[a,b]$ whenever $\esp{\norm{g}_{\EL_p[a,b]}}<\infty$, given $[a,b]\subset\R$.

\section{Setup description}\label{sec:setup}
In this section, we describe the considered WNCS system and state the standing assumptions.
\subsection{Plant and controller}
Consider the continuous-time non-linear plant $\dot{x}_\mathrm{p} = f_\mathrm{p}(x_\mathrm{p},u,w)$, $y = g_\mathrm{p}(x_\mathrm{p})$,
where $x_\mathrm{p}\in\R^{n_\mathrm{p}}$ is the state of the plant, $u\in\R^{n_u}$ is the control input, $w\in\R^{n_w}$ is a deterministic external disturbance which is assumed to have finite $\EL_p$ norm, $y\in\R^{n_y}$ is the plant output, and $n_\mathrm{p},n_u,n_w,n_y\in\N_{>0}$.
As mentioned in the introduction, we follow an emulation approach \cite{wabebu01,nestee04} and thus assume that a stabilising continuous-time controller is designed in absence of the network and has the following non-linear model $\dot{x}_\mathrm{c} = f_\mathrm{c}(x_\mathrm{c},y,w)$, $u = g_\mathrm{c}(x_\mathrm{c})$, where $x_\mathrm{c}\in\R^{n_\mathrm{c}}$ is the controller state, and $n_\mathrm{c}\in\N_{>0}$. When the controller is static, we can simply write $u=g_\mathrm{c}(y)$.
The controller is then implemented over a wireless network so that it no longer receives $y$, but $\hat{y}$, i.e. the vector of the most recently transmitted plant output values which we specify in the next subsection. Similarly, the plant now receives $\hat{u}$ instead of $u$. Therefore, with the wireless network, the plant and controller dynamics are
\begin{subequations}\label{eq:plant-controller}
	\begin{align}
		\dot{x}_\mathrm{p} &= f_\mathrm{p}(x_\mathrm{p},\hat{u},w),\quad y = g_\mathrm{p}(x_\mathrm{p}), \\
		\dot{x}_\mathrm{c} &= f_\mathrm{c}(x_\mathrm{c},\hat{y},w),\quad u = g_\mathrm{c}(x_\mathrm{c}).\label{eq:controller}
	\end{align}
\end{subequations}

The first objective of this work is to establish sufficient conditions on the data transmission rate, for which stability is preserved when the controller is implemented over the network (Section \ref{sec:stability}). The second objective is to use these results to design power control policies that ensure stability of the WNCS (Section \ref{sec:power-control}).

\subsection{Wireless network}
\begin{figure}
	\centering 
	\includegraphics[scale=0.9]{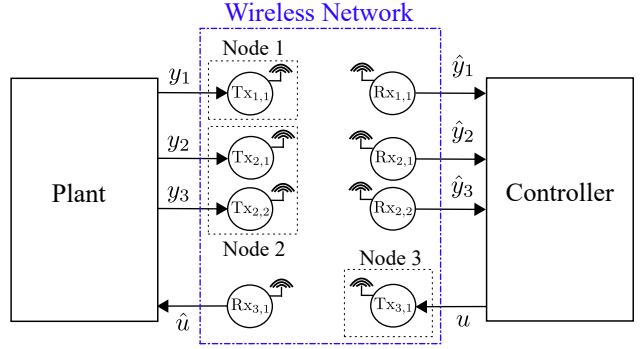}
	\caption{Schematic of a WNCS with three nodes and four links. The network-induced error is given by $e=(e_1,e_2,e_3)$ with $e_1=e_{1,1}=\hat{y}_1-y_1$, $e_2=(e_{2,1},e_{2,2})=(\hat{y}_2-y_2,\hat{y}_3-y_3)$, and $e_3=e_{3,1}=\hat{u}-u$.}
	\label{fig:WNCS}
\end{figure}

A useful change of variable to model and analyse the wireless network is the so-called \emph{network-induced error}, defined as $e\coloneqq  (e_y,e_u)$, where $e_y\coloneqq  \hat{y}-y$ and $e_u\coloneqq  \hat{u}-u$.
We next define the concept of \emph{network node} or \emph{cluster}. A node consists of several sensors and/or actuators (grouped either by their spatial location or merely by convention) with their corresponding data being transmitted at the same transmission instant. Consequently, we partition the network-induced error as $e=(e_1,\dots,e_{N})$, where $e_n$ is the network-induced error associated with the $n$-th node, $n\in\mathcal{N}$, $\mathcal{N}\coloneqq  \{1,\dots,N\}$, and $N\in\{1,\dots,n_y+n_u\}$ is the total number of nodes in the network.
Each node may contain different links (we may also refer to them as channels), thus we write $e_n=(e_{n,1},\dots,e_{n,\ell_n})$, for all $n\in\mathcal{N}$ and associated partition $\ell_n\in\{1,\dots,n_y+n_u\}$. We assume that each link $i_n\in\{1,\dots,\ell_n\}$ within a node $n\in\mathcal{N}$ is associated with a transmitter and receiver denoted by Tx$_{n,i_n}$ and Rx$_{n,i_n}$, respectively. To illustrate our notation more clearly, we refer the reader to Fig. \ref{fig:WNCS} for a wireless network with $N=3$ nodes and $\ell_1+\ell_2+\ell_3=4$ links.

To introduce the wireless network dynamics, we first define the transmission instants, denoted by $\{t_k\}_{k\in\N}$. 
In wireless networks, $\{t_k\}_{k\in\N}$ are often aperiodic and random. In this context, it is common to model them by using renewal processes, and we use a Poisson point process as formalised in the assumption below, see also \cite{hu2019stochastic,agarwal2013exponential,antunes2013stability}.
\begin{assu}\label{assu:transmission-instants}
	Consider a Poisson point process $r(\cdot)$ with rate $\w\in\R_{>0}$ that satisfies $r(t)=0$ for $t\in[0,t_0)$ and $r(t)=k$ for $t\in[t_{k-1},t_k)$, where $\{t_k\}_{k\in\N}$ denotes the sequence of transmission instants defined inductively by: $t_0=\tau_0$ with $\tau_0\sim \text{Exp}(\w)$, and for each $k\in\N_{>0}$, $t_k=t_{k-1}+\tau_k$, with $\tau_k\sim \text{Exp}(\w)$, where the sequence $\{\tau_k\}_{k\in\N}$ is i.i.d.\fin
\end{assu}
The times $\{t_k\}_{k\in\N}$ are also called \emph{arrival times} in the literature, $\{\tau_k\}_{k\in\N}$ are called \emph{intertransmission times} (or \emph{interarrival times}),   $\bar{\tau}\coloneqq 1/\w$ represents the \emph{average intertransmission time}, and $\w$ is the \emph{arrival rate}. Throughout this paper, we will use the terms arrival rate and transmission rate interchangeably.



We next define the random packet loss variables.
\begin{defn}[Packet losses]\label{def:packet-losses}
	Let $f_{n,i_n}$ denote the \emph{probability of successful transmission} in Link $i_n$ within Node $n$. For each node $n\in\mathcal{N}$ we define $\Theta_n(k)=\diag\{\theta_{n,1}(k),\dots,\theta_{n,\ell_n}(k)\}$, for any $k\in\N$, where each $\{\theta_{n,i_n}(k)\}_{k\in\N}$ is a sequence of independent random Bernoulli variables such that $\theta_{n,i_n}(k)=1$ with probability $f_{n,i_n}$, and $\theta_{n,i_n}(k)=0$ with probability $1-f_{n,i_n}$.\fin
\end{defn}
Based on Definition \ref{def:packet-losses}, we say that a transmission is successful in Link $i_n$ if $\theta_{n,i_n}=1$, and the packet is lost whenever $\theta_{n,i_n}=0$.
We assume the following on the packet loss sequences.
\begin{assu}\label{assu:theta}
	$\{\Theta_n(k)\}_{k\in\N}$ and $\{\Theta_m(k)\}_{k\in\N}$ are independent for all $n\neq m$, with $n,m\in\N$, $\{\theta_{n,i_n}(k)\}_{k\in\N}$ and $\{\theta_{n,j_n}(k)\}_{k\in\N}$ are independent for all $i_n\neq j_n$, and  $\{\Theta_n(k)\}_{k\in\N}$ are independent of $\{\tau_k\}_{k\in\N}$.\fin 
\end{assu}
These are standard assumptions in NCS literature that considers Bernoulli packet losses and multiple links, see e.g. \cite{gasigo12}. Different packet loss models are of interest for future work.

Transmissions are governed by a scheduling protocol. Suppose this protocol has determined that Node $n$ attempts to transmit at time instant $t_k$. Then, for a successful transmission ($\theta_{n,i_n}=1$), the corresponding error components $e_{n,i_n}$, $i_n\in\{1,\dots,\ell_n\}$, are set to zero $e_{n,i_n}(t_k^+)=0$, where $e_{n,i_n}(t_k^+)$ is the right limit of $e_{n,i_n}$ at time $t_k$. 
For a packet loss ($\theta_{n,i_n}=0$), the corresponding error components remain unchanged since the signal was not transmitted successfully and thus not updated, $e_{n,i_n}(t_k^+)=e_{n,i_n}(t_k)$.

\section{Stochastic hybrid model for the WNCS}\label{sec:hyrbid-model}
We now formalise the description of the WNCS in Section \ref{sec:setup} into a stochastic hybrid model.
%
We model the WNCS in Fig. \ref{fig:WNCS} as the interconnection of two subsystems, one containing the plant and controller states, and the other the network-induced error state. Let $x\coloneqq  (x_\mathrm{p},x_\mathrm{c})$. The hybrid dynamics of the $x$-subsystem can be obtained directly from \eqref{eq:plant-controller} and the definition of $e$, i.e.,
\begin{subequations}\label{eq:}
	\begin{align}
		\dot{x} &= \mathbf{f}(x,e,w),\ t\in[t_k,t_{k+1}], \\
		x(t_k^+) &= x(t_k),
	\end{align}
\end{subequations}
where $\mathbf{f}(x,e,w) \coloneqq  \big( f_{\rm p}(x_\mathrm{p},e_u+g_\mathrm{c}(x_\mathrm{c}),w) , f_\mathrm{c}(x_\mathrm{c},e_y+g_\mathrm{p}(x_\mathrm{p}),w) \big)$.
Note that since both plant and controller are assumed to be continuous, $x$ remains unchanged after jumps.
On the other hand, the network dynamics are represented via the network-induced error $e$. For the sake of generality, we note that between transmission instants, $\hat{y}$ and $\hat{u}$ are generated according to $\dot{\hat{y}} = \hat{f}_\mathrm{p}(x_\mathrm{p},x_\mathrm{c},\hat{y},\hat{u},w)$ and $\dot{\hat{u}} = \hat{f}_\mathrm{c}(x_\mathrm{p},x_\mathrm{c},\hat{y},\hat{u},w)$ for $t\in[t_k,t_{k+1}]$ and $k\in\N$, where $\hat{f}_\mathrm{p}$ and $\hat{f}_\mathrm{c}$ are defined depending on the implemented processing. For example, the use of zero-order hold devices would lead to $\hat{f}_\mathrm{p}=0$ and $\hat{f}_\mathrm{c}=0$. Therefore, we have
\begin{align}\label{eq:e-dot}
	\dot{e} = (\dot{\hat{y}}-\dot{y},\dot{\hat{u}}-\dot{u}) = \mathbf{g}(x,e,w),\ t\in[t_k,t_{k+1}],
\end{align}
where $\mathbf{g}(x,e,w) \coloneqq  \Big( 
\hat{f}_\mathrm{p}(x_\mathrm{p},x_\mathrm{c},g_\mathrm{p}(x_\mathrm{p})+e_y,g_\mathrm{c}(x_\mathrm{c})+e_u,w)
-\frac{\partial g_\mathrm{p}(x_\mathrm{p})}{\partial x_\mathrm{p}}f_\mathrm{p}(x_\mathrm{p},e_u+g_\mathrm{c}(x_\mathrm{c}),w),
\hat{f}_\mathrm{c}(x_\mathrm{p},x_\mathrm{c},g_\mathrm{p}(x_\mathrm{p})+e_y,g_\mathrm{c}(x_\mathrm{c})+e_u,w)
-\frac{\partial g_\mathrm{c}(x_\mathrm{c})}{\partial x_\mathrm{c}}f_\mathrm{c}(x_\mathrm{c},e_y+g_\mathrm{p}(x_\mathrm{p}),w)
\Big)$.

Next we present the dynamics of the $e$-subsystem at $t_k^+$ with the help of the packet loss sequence $\{\Theta_n(k)\}_{k\in\N}$ from Definition \ref{def:packet-losses}. The informal description of successful/failure packet transmissions below Assumption \ref{assu:theta} can be mathematically written as follows, for any $k\in\N$,
\begin{multline}\label{eq:protocol-loss}
	e(t_k^+) = \diag\{\Theta_1(k),\dots,\Theta_{N}(k)\}h(k,e(t_k)) \\ + (I_{n_e} - \diag\{\Theta_1(k),\dots,\Theta_{N}(k)\})e(t_k),
\end{multline}
where $h$ is a mapping that determines which node attempts to transmit.
The mathematical description of $h$ depends on the implemented protocol, see e.g. \cite{nestee04}. We illustrate this fact in Section \ref{sec:protocol}. 

With all the above, we can write the following stochastic hybrid model for the WNCS in Fig. \ref{fig:WNCS},
\begin{subequations}\label{eq:WNCS}
	\begin{align}
		\dot{x} &= \mathbf{f}(x,e,w),\hspace{1.4mm} t\in[t_k,t_{k+1}]\label{eq:x-sys}\\
		\dot{e} &= \mathbf{g}(x,e,w),\hspace{1mm} t\in[t_k,t_{k+1}]\label{eq:e-sys} \\
		x(t_k^+) &= x(t_k),\label{eq:x-sys2} \\
		e(t_k^+) &= \mathbf{h}(k,e(t_k)), \label{eq:protocol}
	\end{align}
\end{subequations}
where $\mathbf{h}(k,e)\coloneqq  \diag\{\Theta_1(k),\dots,\Theta_{N}(k)\}h(k,e) + (I_{n_e} - \diag\{\Theta_1(k),\dots,\Theta_{N}(k)\})e$. We emphasise that it is only the network-induced constraints (random transmissions and packet losses) that introduce randomness in \eqref{eq:WNCS}, and the disturbance $w$ is an $\EL_p$ signal as per Section \ref{sec:setup}. This leads to randomness in the jump equation \eqref{eq:protocol} and the transmission instants $t_k$. For detailed information about construction of solutions to the stochastic hybrid system (SHS) \eqref{eq:WNCS}, we refer the reader to \cite{tabnes08-tac}, see also \cite{hespanha2005model,teel2014stability}. At a general level, we flow the continuous dynamics until a discrete event occurs, and then repeat from the new state after the jump. Note that Standing Assumption \ref{assu:transmission-instants} implies $\Pr\{t_{k+1}-t_k=0\}=0$, which ensures that a bounded 	number of transmissions (jumps) occur in any bounded time interval with probability one (Zeno solutions are a.s. not possible).


It is worth noticing that the SHS \eqref{eq:WNCS} covers a broader scenario than the one adopted in previous works on stochastic non-linear WNCS such as \cite{wang2011quantized,li2018input}, where only classes of non-linear systems that are input-affine were considered.
In the context of general non-linear systems, model \eqref{eq:WNCS} generalises the network-induced error dynamics $e(t_k^+)=\theta(k)h(k,e(t_k))+(1-\theta(k))e(t_k)$ considered in \cite{tabnes08-tac,hestee06}, where $\{\theta(k)\}_{k\in\N}$ is an i.i.d. sequence of Bernoulli random variables such that $\theta(k)=1$ with probability $f$ and $\theta(k)=0$ with probability $1-f$. When applied to a context of multiple wireless links, \cite{tabnes08-tac} would assume the probability of success to be the \emph{same} for every link and every node, which is often unrealistic. On the contrary, we consider that every link in the network has its own success probability, which is tailored to more practical wireless scenarios, see e.g. IIoT applications like intelligent transportation systems \cite{dey2016vehicle} or multi-hop industrial networks such as WirelessHART \cite{chen2010wirelesshart}.

\section{Underlying scheduling protocols}\label{sec:protocol}
We consider two types of scheduling mechanisms that cover a range of applications in wireless communications. These are \emph{stochastic protocols} and \emph{deterministic protocols}, which have been extensively considered in the NCS literature in different contexts, see e.g. \cite{nestee04,tabnes08-tac}. 

\subsection{Stochastic protocols}\label{sec:protocol-stochastic}
Here, the map $h$ in \eqref{eq:protocol-loss} randomly chooses a node/cluster to transmit.
This protocol has been widely studied in the literature, see e.g. \cite{sheng2019distributed,zou2017mathcal,donkers2012stability}.
Formally, we define $h$ in \eqref{eq:protocol-loss} as $h(k,e) = \mathcal{H}(k) e$, for any $k\in\N$ and $e\in\R^{n_e}$, where $\{\mathcal{H}(k)\}_{k\in\N}$ are i.i.d. random matrices taking values in the finite set $\mathbb{M}\coloneqq  \{ M_1,\dots,M_N \}$, where each $M_n$, $n\in\mathcal{N}$, is such that
$M_n e = (e_1,\dots,e_{n-1},0,e_{n+1},\dots,e_N)$, for any $e\in\R^{n_e}$.
The set $\mathbb{M}$ contains all the possible matrices that result from each contending node getting access to transmit. At every transmission instant $t_k$, $\mathcal{H}(k)$ will be equal to some $M_n$, i.e. some node $n$ will be granted network access. However, packets can still be lost and thus the dynamics in \eqref{eq:protocol-loss}, become, for any $k\in\N$
\begin{align}\label{eq:protocol-stochastic}
	e(t_k^+) = Q(k)e(t_k),
\end{align}	
where $Q(k) \coloneqq   \diag\{\Theta_1(k),\dots,\Theta_{N}(k)\}(\mathcal{H}(k)-I_{n_e})+I_{n_e}$. The probability that Node $n$ transmits successfully through the network is given by $\Pr\{Q(k) = M_n \} = f_n/N$,
where $f_n\coloneqq  f_{n,1}\cdots f_{n,\ell_n}$ is the cumulative successful probability of Node $n$. Recall that $f_{n,i_n}$, $i_n\in\{1,\dots,\ell_n\}$, comes from Definition \ref{def:packet-losses}.

For deterministic networks, the authors in \cite{tanete07} introduced the notion of persistently exciting (PE) protocols.
Essentially, a PE protocol would visit every network node at least once in a finite number of transmissions. This notion was later extended to the stochastic case in \cite{tabnes08-tac} and named \emph{almost surely (a.s.) covering protocol.}
Before defining it, we introduce two preliminary definitions, \emph{cover times} and \emph{covering sequence}.

\begin{defn}[Cover times]\label{def:cover-time}
	For any $k\in\N$, the $k$-th \emph{cover time}, denoted by $T_k$, is defined as 
	$T_k \coloneqq  \min\big\{ j\geq 1 :  \{M_1,\dots,M_N\}\subset \{Q(T_{k-1}),\dots,Q(T_{k-1}+j-1)\}  \big\}$,
	and $T_{-1}=0$. \fin
\end{defn}
Collectively, $\{T_k\}_{k\in\N}$ is referred to as the cover time process. Cover times are the times in which every network node has been visited at least once. This notion is closely related to the cover time of an undirected graph, see e.g. \cite{chandra1996electrical}, and the Coupon's Collector problem \cite{upfal2005probability}.

\begin{defn}[Covering sequence]
	Let $\tau_k = t_k - t_{k-1}$ satisfy Assumption \ref{assu:transmission-instants}. We say that $C(j,k)\coloneqq  \{(Q(j),\tau_j),\dots,(Q(k),\tau_k)\}$ is a \emph{covering sequence} if and only if for any $j\in\N$ there exists $k\geq j$ such that 	
	$\{M_1,\dots,M_N\}\subset \{Q(j),\dots,Q(k)\}$.\fin
\end{defn}
We note that cover times are the lengths of consecutive disjoint covering sequences.
We can now state the notion of a.s. covering protocols introduced in \cite{tabnes08-tac}.

\begin{defn}[A.s. covering protocol]\label{def:covering}
	A stochastic protocol is \emph{a.s. covering} if $\Pr\{T_k<\infty\}=1$, $\forall k\geq 0$, with $T_k$ as per Definition \ref{def:cover-time}.\fin
\end{defn}

An a.s. covering protocol grants network access to every node---at least once---within a finite number of transmissions with probability one.
The stochastic protocol \eqref{eq:protocol-stochastic} considered in this paper is a.s. covering provided the probabilities of success are not equal to zero.

To show stability under stochastic protocols, properties of the cover times $T_k$ are provided in the lemma below.

\begin{lemma}\label{lem:pgf-T}
	Let $T$ be the cover time for the sequence $\{(Q(0),\tau_0),\dots,(Q(T-1),\tau_{T-1})\}$.  Then, 
	\begin{align}
		\esp{T} &= \sum_{n=1}^{N} \frac{N}{[N-(n-1)]f_n}, \label{eq:ET}\\
		G_T(s) &= \prod_{n=1}^{N} \frac{ s(N-(n-1))f_n }{ N(1-(1-f_n)s) - s(n-1)f_n },\label{eq:G_T}
	\end{align}
	for any $s\in\R$, with $|s| < \frac{1}{1-\left[\min_{n\in\mathcal{N}} f_n(N-n+1)\right]/N}$.
\end{lemma}
\textbf{PROOF:}
Let $n\in\mathcal{N}$, and $\bar{t}_n$ be the additional number of transmissions required to go from $n-1$ to $n$ different nodes being covered. 
Therefore, $\bar{t}_n$ is geometrically distributed with parameter $[N-(n-1)]f_n/N$,
and $T$ can be written as $T = \sum_{n=1}^{N} \bar{t}_n$. Then, by linearity of expectation and since $\{Q(k)\}_{k\in\N}$ are i.i.d. according to Assumption \ref{assu:theta}, we have $\esp{T} = \espp{\sum_{n=1}^{N}\bar{t}_n} = \sum_{n=1}^{N}\esp{\bar{t}_n}= \sum_{n=1}^{N} N/((N-(n-1))f_n)$, which corresponds to \eqref{eq:ET}.
We now prove \eqref{eq:G_T}, $G_T(s) \coloneqq  \esp{s^T} = \espp{s^{(\bar{t}_1+\cdots + \bar{t}_N)}} = \prod_{n=1}^{N} \espp{s^{\bar{t}_n}}$, for any $s\in\R$, where $\espp{s^{\bar{t}_n}}$ corresponds to the p.g.f. $G_{\bar{t}_n}(s)$. Since $\bar{t}_n$ is geometric, it is well known that
$G_{\bar{t}_n}(s) = \frac{s\bar{p}}{1-s(1-\bar{p})}$ \cite{stir05}, for $|s|<1/(1-\bar{p})$,
where $\bar{p}$ is the parameter of the geometric variable $\bar{t}_n$, which is $\bar{p}=[N-(n-1)]f_n/N$ as stated above. With the latter, \eqref{eq:G_T} follows immediately.\qedd

We note that $\esp{T}$ represents the expected number of transmissions in which all matrices $M_1,\dots,M_N$ have been covered at least once. We see in \eqref{eq:ET} that it depends on the cumulative probability of success $f_n$ of each node $n\in\mathcal{N}$. This differs from previous works like \cite{tabnes08-tac} in which $\espp{T}$ only depends on a single probability of success.


\subsection{Deterministic protocols}\label{sec:protocol-uges}
Deterministic protocols schedule network nodes based on predetermined rules. For example, 
protocols such as round robin (RR) fit this framework \cite{wheels93}.
Mathematically, we consider static deterministic protocols for which $h$ in \eqref{eq:protocol-loss} takes the form $h(k,e(t_k)) = \mathcal{H}(k)e(t_k)$, where $\mathcal{H}(k)$ is a deterministic diagonal matrix with elements taking values in $\{0,1\}$ depending on the predetermined scheduling rule. We illustrate this with an example further below.
We also note that, even though the scheduling mechanism is deterministic, the jump map \eqref{eq:protocol} is still stochastic.

Below, we define an important property, firstly introduced in \cite{tabnes08-tac}.

\begin{defn}[a.s. UGES protocols]\label{def:as-UGES}
	We say that \eqref{eq:protocol} is a.s. UGES with Lyapunov function $W:\N\times\R^{n_e}\rightarrow \R_{\geq 0}$ if there exist a sequence of independent random variables $\kappa_k\in\R_{\geq 0}$ and $a_1,a_2,\bar{\kappa}\in\R_{>0}$ such that the following conditions hold for the auxiliary discrete-time system $e(k+1)=\mathbf{h}(k,e(k))$
	\begin{subequations}
		\begin{align}
			a_1|e| &\leq W(k,e) \leq a_2|e|,\label{eq:W-a1a2} \\
			W(k+1,\mathbf{h}(k,e)) &\leq \kappa_kW(k,e)  ,\label{eq:W-kappa}\\
			\espp{\kappa_k} &\leq \bar{\kappa} < 1,  \label{eq:W-esp-kappa}
		\end{align}
	\end{subequations}
	for all $k\in\N$ and all $e\in\R^{n_e}$.\fin
\end{defn}

A.s. UGES protocols are more general than the well-known notion of UGES introduced in \cite{nestee04} for deterministic networks, where $\bar{\kappa}=\eta$ for some $0\leq \eta <1$. Particularly, Definition \ref{def:as-UGES} allows to account for packet losses through the stochastic process $\kappa_k$. Finding $a_1,a_2,\kappa_k$ and $\bar{\kappa}$ in Definition \ref{def:as-UGES} is done case-by-case. We present an example at the end of this section.

We now provide a result stating that any UGES protocol is also a.s. UGES under some conditions (proof is omitted due to space constraints).
\begin{lemma}\label{lem:UGES-is-as-UGES}
	Suppose that \eqref{eq:protocol} is UGES in the sense of \cite{nestee04} on a lossless network (i.e. $\Theta_n(k)=I$ for all $n\in\{1,\dots,N\}$ and $k\in\N$). That is, there exist $W:\N\times \R^{n_e}\rightarrow \R_{\geq 0}$,  $a_1,a_2\in\R_{>0}$, and $0\leq \eta < 1$ such that the following holds for the auxiliary discrete-time system $e(k+1)=h(k,e)$,
	\begin{subequations}\label{eq:UGES}
		\begin{align}
			a_1|e| &\leq W(k,e) \leq a_2|e|, \label{eq:W-bounds-uges}\\
			W(k+1,h(k,e)) &\leq \eta W(k,e), \label{eq:W+-uges}
		\end{align} 
	\end{subequations}
	for all $k\in\N$ and all $e\in\R^{n_e}$.
	Then, \eqref{eq:protocol} is a.s. UGES as per Definition \ref{def:as-UGES} if $P_{\eta}(k)\eta + 1 - P_{\eta}(k) \leq \bar{\kappa} < 1$, where $P_{\eta}(k) \coloneqq  \Pr\{\kappa_k=\eta\}$.\qedd 
\end{lemma}

We now depict how to show that a protocol is a.s. UGES. For illustrative purposes, we pick a round robin protocol, but any other static protocol follows a similar procedure. For round robin, $h(k,e)=\mathcal{H}(k)e$, where $\mathcal{H}(k)\coloneqq  I_{n_e} - \Psi(k)$,
with $\Psi(k)\coloneqq \diag\{\psi_1(k)I_{s_1},\dots,\psi_N(k)I_{s_N}\}$, $\sum_{n=1}^{N}s_n=n_e$, and $\psi_n(k) =1$ if $k = n+\sigma N,\ \sigma\in\N$, or $\psi_n(k) =0$ otherwise, 
where $\psi_n$ is the variable in charge of visiting the nodes one by one. 

Next, we show that the RR protocol is a.s. UGES.

\begin{lemma}\label{lem:RR}
	The round robin protocol satisfies Definition \ref{def:as-UGES} with Lyapunov function $W(k,e) = \sqrt{\sum_{j=k}^{\infty} |\phi(j,k,e)|^2 }$,
	where $\phi(j,k,e)$ is the solution to the auxiliary discrete-time system $e(k+1)=\mathcal{H}(k)e(k)$ at time $j$, starting at time $k$ and initial condition $e$,  $a_1=1$, $a_2=\sqrt{N}$, and with $\kappa_k$ and $\bar{\kappa}$ given by
	\begin{align}
		\kappa_k &= \left( \textstyle \sum_{n=1}^{N} \left( \prod_{i_n=1}^{\ell_n} \theta_{n,i_n}(k) \right) \psi_n(k) \right) \eta \nonumber \\ 
		&\qquad + 1 - \left(\textstyle \sum_{n=1}^{N} \left( \prod_{i_n=1}^{\ell_n} \theta_{n,i_n}(k) \right)\psi_n(k) \right), \label{eq:kappa-RR} \\
		\bar{\kappa} &= 1 - \big( \min_{n\in\mathcal{N}} f_{n,1}\cdots f_{n,\ell_n} \big) (1-\eta). \label{eq:mu}
	\end{align}
\end{lemma}
\textbf{PROOF:}
From \cite[Proposition 4]{nestee04} we know that the RR protocol is UGES with the Lyapunov function $W$ as per the statement. Also, $a_1=1,a_2=\sqrt{N}$, and $\eta = \sqrt{N-1/N}$ in \eqref{eq:UGES}. Since RR is UGES, it only remains to use Lemma \ref{lem:UGES-is-as-UGES} to show that it is also a.s. UGES.
Note that $\kappa_k\in\{\eta,1\}$. Then, the expression of $\kappa_k$ follows from using the definition of $\{\Theta_n(k)\}_{n=1,\dots,N}$. This reads as follows. Given the definition of $\psi_n(k)$, only one element of the summation in \eqref{eq:kappa-RR} is active at transmission time $t_k$. For instance, Node $n=1$ transmits at time instant $t_1$ and thus $\kappa_1 = \big(\textstyle \prod_{i_1=1}^{\ell_1} \theta_{1,i_1}(1) \big) \eta + 1 - \big(\textstyle \prod_{i_1=1}^{\ell_1} \theta_{1,i_1}(1) \big)$.
If every link in Node 1 had successful transmissions at time instant $t_1$, then $\theta_{1,1}(1)=\cdots = \theta_{1,\ell_1}(1)=1$ and $\kappa_1=\eta$. In any other case $\kappa_1=1$ as expected. This is generalised for every time instant in \eqref{eq:kappa-RR}.
We now compute the corresponding $P_{\eta}(k)=\Pr\{\kappa_k=\eta\}$ by using \eqref{eq:kappa-RR}, that is, $P_{\eta}(k) = \sum_{n=1}^{N} \Pr\{\theta_{n,1}(k)=\cdots=\theta_{n,\ell_n}(k)=1 \} \psi_n(k) = \sum_{n=1}^{N} \left( f_{n,1}\cdots f_{n,\ell_n} \right) \psi_n(k)$,
from which the bound $\bar{\kappa}$ in \eqref{eq:mu} follows. Lastly, we note that $\bar{\kappa}<1$ provided the success probabilities are non-zero, and thus by Lemma \ref{lem:UGES-is-as-UGES} we conclude that the RR protocol is a.s. UGES.\qedd



\section{Stability results}\label{sec:stability}
This section provides the main stability results that relate channel probabilities to transmission rate, which serve as a foundation for the upcoming transmit power control policies. To cope with the plant external disturbances, the primary notion we consider in this paper is the stochastic counterpart of the classical input-output $\EL_p$ stability, which is called \emph{$\EL_p$ stability in expectation}. 
It has been adopted in the context of WNCS in, e.g., \cite{tabnes08-tac,li2019string}. 

\begin{defn}[$\EL_p$ stability in expectation]\label{def:Lp-in-exp}
	For system \eqref{eq:WNCS}, define $z\coloneqq  (x,e)$ and consider any input and output functions $\mathscr{U}=F(z,w)$, $\mathscr{Y}=H(z,w)$, respectively.
	Let $p\in\N\cup\{+\infty\}$ and $\gamma\geq 0$ be given. We say that \eqref{eq:WNCS} is \emph{$\EL_p$ stable in expectation} from $\mathscr{U}$ to $\mathscr{Y}$ with gain $\gamma$ if there exists $K\geq 0$ such that any solution to \eqref{eq:WNCS} with input $\mathscr{U}\in\EL_p^e$ verifies $\espp{\norm{\mathscr{Y}}_{\EL_p[0,t]}} \leq K|z(0)| + \gamma \espp{\norm{\mathscr{U}}_{\EL_p[0,t]}}$
	for all $t\geq 0$.\fin
\end{defn}
We highlight again that the SHS \eqref{eq:WNCS} has stochastic jumps but deterministic flow. What introduces randomness in the jumps are packet losses and transmissions; hence still leading to random solutions. However, the external disturbance $w$ in \eqref{eq:WNCS} is a deterministic $\EL_p$ signal and not a Levy process as in other WNCS literature like \cite{hestee06}. Note that we are concerned with robustness of stability in the sense of, e.g., \cite{dulpag99}, for which Definition \ref{def:Lp-in-exp} is more suitable. Disturbances modelled by Levy processes may be more appropriate in other contexts such as modelling sensor noise for instance.
In absence of plant disturbances, i.e., for $w=0$, we will show additional stability properties such as exponential stability (in expectation and probability), see Section \ref{sec:stability-more}.

\subsection{Stability under stochastic scheduling}\label{sec:stability-stoch}
When the WNCS \eqref{eq:WNCS} implements the a.s. covering stochastic protocol from Section \ref{sec:protocol-stochastic}, we show that stability of the WNCS is guaranteed with sufficiently frequent data transmission and some assumptions on \eqref{eq:WNCS}, which is formalised in the below theorem.

\begin{theorem}\label{theo:stability-stoch-protocol}
	Consider the WNCS \eqref{eq:WNCS} and suppose the following holds.
	\begin{enumerate}[$(i)$]
		\item The scheduling protocol is stochastic and a.s. covering as in Section \ref{sec:protocol-stochastic}. 
		\item The $e$--dynamics in \eqref{eq:e-sys} satisfy
		\begin{align}\label{eq:g-dynamics}
			\overline{\mathbf{g}}(x,e,w) \preceq A \overline{e} + \tilde{y}(x,w),
		\end{align}
		for all $(x,e,w)\in \R^{n_x}\times \R^{n_e}\times \R^{n_w}$ and $k\in\N$, where $A$ is a symmetric $n_e\times n_e$ matrix with non-negative entries, and $\tilde{y}(x,w)=G(x)+E\overline{w}$ for some $G:\R^{n_x}\rightarrow \R^{n_e}$ and $E\in\R^{n_e\times n_w}$.
		\item The $x$-subsystem \eqref{eq:x-sys},\eqref{eq:x-sys2} is $\EL_p$ stable in expectation from $(e,w)$ to $G(x)$ with finite gain $\gamma$ for some $p\in \N\cup \{+\infty\}$.
	\end{enumerate}
	Then, there exists an arrival rate $\w\in(0,\infty)$ that satisfies $\w > N|A|/[\min_{n\in\mathcal{N}} f_n(N-n+1)]$, and
	\begin{align}\label{eq:small-gain-condition}
		\left(\sum_{n=1}^{N} \frac{N}{[N-(n-1)]f_n}\right)\frac{ \gamma(1+\rho_{\w}) }{ (\w - |A|)(1 - \rho_{\w}) } < 1,
	\end{align}
	such that the WNCS \eqref{eq:WNCS} is $\EL_p$ stable in expectation from $w$ to $(G(x),e)$ with finite linear gain, where
	\begin{align}\label{eq:rho}
		\rho_{\w} = \prod_{n=1}^{N} \frac{ (N-(n-1))f_n }{ N(1-|A|/\w-(1-f_n)) - (n-1)f_n }-1.
	\end{align}
\end{theorem}
\textbf{PROOF:}
	See Appendix \ref{sec:proof-stab-stoch-prot}.\qedd

Condition $(ii)$ is the standard vector analogue of the dissipation-type inequality \eqref{eq:dWdt} often adopted in the NCS literature, see e.g. \cite{tabnes08-tac,nestee04}. 
Condition $(iii)$ is ensured when designing the controller in the first step of emulation, we show how to systematically verify this condition for LTI systems in Section \ref{sec:linear}. Our results are based on small-gain arguments, and Theorem \ref{theo:stability-stoch-protocol} ensures stability of the WNCS \eqref{eq:WNCS} provided the transmission rate is chosen to satisfy the small-gain condition \eqref{eq:small-gain-condition}. Even though this is not a closed-form bound on $\w$ given the generality introduced by the multiple links and nodes, it can be easily solved numerically.
We can see from \eqref{eq:small-gain-condition} that the different success probabilities and the number of nodes play an important role in the stability bound on $\w$.

\subsection{Stability under deterministic scheduling}\label{sec:stability-det}
Here we present the corresponding stability result for deterministic protocols satisfying the a.s. UGES property in Definition \ref{def:as-UGES}. 
\begin{theorem}\label{theo:Lp-overall}
	Consider the WNCS \eqref{eq:WNCS} and suppose the following holds.
	\begin{enumerate}[$(i)$]
		\item \eqref{eq:protocol} is a.s. UGES with Lyapunov function $W$.
		%
		\item There exists $L\geq 0$ such that
		\begin{align}\label{eq:dWdt}
			\left\langle \partial W/\partial e,\mathbf{g}(x,e,w)\right\rangle \leq LW(k,e) + |\tilde{y}(x,w)|
		\end{align}
		holds for almost all $e\in\R^{n_e}$, all $(x,w)\in\R^{n_x}\times\R^{n_w}$ and $k\in\N$, with  $\tilde{y}= G(x)+ Ew$ for some  $G:\R^{n_x}\rightarrow \R^{n_e}$ and $E\in\R^{n_e\times n_w}$.
		\item The $x$-subsystem  \eqref{eq:x-sys},\eqref{eq:x-sys2} is $\mathcal{L}_p$ stable from $(W(e),w)$ to $\tilde{y}(x,w)$ with finite  gain $\gamma$ for some $p\in\N\cup\{+\infty\}$.
	\end{enumerate}
	Then, there exists an arrival rate $\w\in(0,\infty)$ that satisfies $\w>L/(1-\bar{\kappa})$ and
	\begin{align}\label{eq:omega-numeric}
		\gamma s_{\infty}(\w)/(\w - L) < 1,
	\end{align}
	such that the WNCS \eqref{eq:WNCS} is $\EL_p$ stable in expectation from $w$ to $(G(x),W(e))$ with finite linear gain, where $s_{\infty}(\w)\coloneqq \textstyle  \sum_{j=0}^{\infty} \big[  \big(\frac{\w}{\w-L}\big)^j\times \prod_{\iota=0}^{j-1}\esp{\kappa_\iota}\big]<\infty$.
\end{theorem}
\textbf{PROOF:}
	 See Appendix \ref{sec:appendix-stability-det}.\qedd

Condition $(ii)$ is an exponential growth assumption on the $e$-subsystem. It is satisfied when $W$ is globally Lipschitz in $e$ uniformly in $k$, which is the case by many protocols, see e.g. \cite{povane14,nestee04}, and when $\mathbf{g}$ can be linearly bounded from above for instance.
Condition $(iii)$ can be ensured when designing the controller in the first step of emulation. Any stabilisable and detectable linear time-invariant system does ensure this property, see Section \ref{sec:linear} further below. For examples of non-linear systems that satisfy this condition we refer the reader to e.g. \cite{neteca09,abpoda16}.

Similar to the stochastic protocol case (cf. Theorem \ref{theo:stability-stoch-protocol}), the success probabilities also play a role in stability. In this case, they do it through $\espp{\kappa_k}$, which depends on these probabilities as we showed in Lemma \ref{lem:RR} for the round robin protocol.
Also, the bound on $\w$ that comes from the small-gain condition \eqref{eq:omega-numeric} cannot be written in a closed-form, but it can be easily computed numerically.

We next show that the transmission rate bounds from Theorems \ref{theo:stability-stoch-protocol} and \ref{theo:Lp-overall} are less conservative than the ones in \cite{tabnes08-tac}. Consider the framework from \cite{tabnes08-tac}, where all the wireless links have a unique probability of success which we denote by $f$. Note that our model \eqref{eq:WNCS} reduces to \cite{tabnes08-tac} when $f_{n,i_n}=f$ for all $n\in\mathcal{N}$ and $i_n\in\{1,\dots,\ell_n\}$. Denote the transmission rate bound from \cite{tabnes08-tac} as $\w_{\text{\tiny \cite{tabnes08-tac}}}$, and the bounds provided in Theorems \ref{theo:stability-stoch-protocol} and \ref{theo:Lp-overall} by $\w_{\text{\tiny Thm.\ref{theo:stability-stoch-protocol}}}$ and $\w_{\text{\tiny Thm.\ref{theo:Lp-overall}}}$, respectively. 
\begin{lemma}\label{lem:tabbara}
	Theorem 6.1 in \cite{tabnes08-tac} (resp. Theorem 7.1) ensures $\EL_p$ stability in expectation of the WNCS \eqref{eq:WNCS} only if $f=\min_{n\in\mathcal{N}}f_n$. Moreover, $\w_{\text{\tiny Thm.\ref{theo:stability-stoch-protocol}}}\leq \w_{\text{\tiny \cite[Thm.6.1]{tabnes08-tac}}}$ (resp. $\w_{\text{\tiny Thm.\ref{theo:Lp-overall}}}\leq \w_{\text{\tiny \cite[Thm.7.1]{tabnes08-tac}}}$), where equality only holds when $f_n=f$ for all $n\in\mathcal{N}$.
\end{lemma}
\textbf{PROOF:} See Appendix \ref{sec:proof-Tabbara}.\qedd

Lemma \ref{lem:tabbara} states that the stability results in \cite{tabnes08-tac} are only applicable to our setup if their probability of transmission is equal to the worst cumulative probability between the nodes, and thus the resulting stability bounds in \cite{tabnes08-tac} will always be larger (worse) than the ones we provide. In the trivial case $f_n=f$, $n\in\mathcal{N}$, our bounds recover the ones in \cite{tabnes08-tac}. It is not straightforward to analytically compute the distance between the bounds, but we illustrate in the example in Section \ref{sec:numerical} that our bounds can be significantly less conservative (around 40-50\% of improvement).
The practical impact of Lemma \ref{lem:tabbara} is that more conservative bounds, such as the ones in \cite{tabnes08-tac}, translate into requiring faster transmissions in the network, which might not be feasible with the available hardware.

\subsection{Further stability properties}\label{sec:stability-more}
We now state additional stability properties.
The next corollary follows from Theorem \ref{theo:stability-stoch-protocol}.
\begin{corollary}
	Suppose that for some $p\in\N\cup\{+\infty\}$, the following holds.
	\begin{enumerate}[$(i)$]
		\item The WNCS \eqref{eq:WNCS} under stochastic protocols is $\EL_p$ stable in expectation from $w$ to $(G(x),e)$.
		\item System \eqref{eq:WNCS} is $\EL_p$ to $\EL_p$ detectable in expectation from $(G(x),e)$, i.e.
		there exists $\tilde{K}\geq 0$ and $\tilde{\gamma}\geq 0$ such that any solution to \eqref{eq:WNCS} with input $w\in\EL_p$ verifies
		$\espp{\norm{(x,e)}_{\EL_p[0,t]}} \leq \tilde{K}|(x(0),e(0))| + \tilde{\gamma} \espp{\norm{(G(x),e)}_{\EL_p[0,t]}} + \tilde{\gamma} \norm{w}_{\EL_p[0,t]}$, for all $t\geq 0$.
	\end{enumerate}
	Then, the WNCS \eqref{eq:WNCS} is $\EL_p$ stable in expectation from $w$ to $(x,e)$.\qedd 
\end{corollary}

Analogously, for deterministic protocols (c.f. Theorem \ref{theo:Lp-overall}), if the WNCS \eqref{eq:WNCS} is $\EL_p$ stable in expectation from $w$ to $(G(x),W(e))$, and $\EL_p$ to $\EL_p$ detectable in expectation from $(G(x),W(e))$, then the WNCS \eqref{eq:WNCS} is $\EL_p$ stable in expectation from $w$ to $(x,e)$.

It is also important to state stability properties in the absence of disturbances (i.e. $w=0$ in \eqref{eq:WNCS}). In this context, consider the following definition; see \cite[Section 4]{teel2014stability} for similar stability concepts regarding SHSs.

\begin{defn}
	Consider system \eqref{eq:WNCS} and let $w=0$. We say that \eqref{eq:WNCS} is \emph{uniformly globally exponentially stable (UGES)} in expectation if, for all $(x(0),e(0))\in\R^{n_x}\times\R^{n_e}$ and corresponding solution, 	
	there exist $K> 0$ and $c> 0$ such that $\espp{|(x(t),e(t))|}\leq K\exp(-ct)|(x(0),e(0))|$ for all $t\geq 0$.\fin
\end{defn}

A slight extension of \cite[Theorem 3]{nestee04} allows us to state exponential stability in expectation.
\begin{corollary}\label{coro:GES}
	Suppose that for some $p\in\N\cup\{+\infty\}$, the WNCS \eqref{eq:WNCS} is $\EL_p$ stable in expectation from $w$ to $(x,e)$, and  
	\begin{align}
		(\exists L_1\geq 0)\quad |\mathbf{f}(x,e,0)| &\leq L_1 (|x|+|e|),  \label{eq:L1}\\
		(\exists L_2\geq 0)\quad |\mathbf{g}(x,e,0)| &\leq L_2(|x|+|e|), \label{eq:L2}\\
		(\exists L_3\geq 0)\quad |\mathbf{h}(k,e)| &\leq L_3 |e|, \label{eq:L3} 
	\end{align}
	for all $x\in \R^{n_x}$, $e\in\R^{n_e}$, and $k\in\N$. Then, the WNCS \eqref{eq:WNCS} with $w=0$ is UGES in expectation.\qedd 
\end{corollary}

\begin{remark}
	Let $\Theta(k)\coloneqq  \diag\{\Theta_1(k),\dots,\Theta_N(k)\}$ and recall that $\mathbf{h}(k,e)=\Theta(k)h(k,e)+(I-\Theta(k))e$ (c.f. \eqref{eq:protocol}). Moreover, recall that $h(k,e)=\mathcal{H}(k)e$ where $\mathcal{H}(k)$ is a diagonal matrix with elements taking values in the set $\{0,1\}$ either deterministically or stochastically depending on the underlying protocol. Therefore, from the definition of $\Theta(k)$ (c.f. Definition \ref{def:packet-losses}), it is immediate to note that \eqref{eq:L3} holds with $L_3=1$, since $|\mathbf{h}(k,e)|\leq |\Theta(k)(\mathcal{H}(k)-I_{n_e})+I_{n_e}||e|\leq |e|$.
	We emphasise that for LTI systems, \eqref{eq:L1} and \eqref{eq:L2} always hold  and thus UGES in expectation can be stated directly from $\EL_p$ stability in expectation. We detail this fact in Section \ref{sec:linear}.\fin 
\end{remark}

Additionally, we can derive exponential stability in probability (proof is omitted due to space constraints).
\begin{corollary}\label{coro:GES-in-prob}
	Under the conditions of Corollary \ref{coro:GES} (i.e.  \eqref{eq:WNCS} is UGES in expectation), then \eqref{eq:WNCS} is UGES in probability, that is, for any $\epsilon\in(0,1)$, there exist $\mathscr{K}>0$ and $c>0$ such that $\Pr\big\{|(x(t),e(t))|\leq \mathscr{K} e^{-ct} |(x(0),e(0))| \big\} 
	\geq 1 - \epsilon$,
	for all $t\geq 0$, $(x(0),e(0))\in\R^{n_x}\times\R^{n_e}$, and corresponding solution $(x(\cdot),e(\cdot))$.\qedd 
\end{corollary}

For a general non-linear setting such as the one considered in this paper, stating additional properties like \emph{mean-square stability} (MSS) is not straightforward. However, MSS may be stated provided the functions $W$ and $G$ satisfy extra requirements, similarly to \cite{hestee06}. That is, if, for instance, $\exists \alpha,\beta>0$ such that $G(x)\geq \alpha |x|^2$ and $W(k,e)\geq \beta|e|^2$, $\forall x\in\R^{n_x}, k\in\N,e\in\R^{n_e}$.

\begin{figure*}[!htp]
	\hrule
	\begin{align}
		\smallmtx
		\begin{bmatrix}
			A_{11}^\top \textbf{P} + \textbf{P}A_{11} + {A}_{21}^\top {A}_{21} + \bm{\varepsilon}_1 I_{n_x} & \textbf{P} A_{12} & \textbf{P}E_1  \\
			A_{12}^\top \textbf{P} & \bm{\varepsilon}_2 I_{n_e} - \bm{\mu} I_{n_e} & 0 \\
			E_1^\top \textbf{P} & 0 & -\bm{\mu} I_{n_w}
		\end{bmatrix}
		&\leq 0, \label{eq:LMI-stoch} \\
		\smallmtx
		\begin{bmatrix}
			A_{11}^\top \textbf{Q} + \textbf{Q}A_{11} + L_1^2A_{21}^\top A_{21} + \bm{\varepsilon}_3 I_{n_x} & \textbf{Q} A_{12} & \textbf{Q}E_1 + L_1^2 A_{21}^\top E_2 \\
			A_{12}^\top \textbf{Q} & \bm{\varepsilon}_4 I_{n_e} - \bm{\vartheta} I_{n_e} & 0 \\
			L_1^2E_2^\top A_{21} + E_1^\top \textbf{Q} & 0 & L_1^2 E_2^\top E_2 -\bm{\vartheta}/a_1^2 I_{n_w}
		\end{bmatrix}
		&\leq 0. \label{eq:LMI}
	\end{align}
	\hrule
\end{figure*}

\section{Linear time-invariant systems}\label{sec:linear}
In this section, we establish stability results for LTI systems by showing how to systematically verify the conditions used in Section \ref{sec:stability}. Particularly, we tailor  Theorem \ref{theo:stability-stoch-protocol} and Theorem \ref{theo:Lp-overall} to LTI systems in terms of LMI conditions. We focus on $\EL_2$ stabilisation. Consider the following LTI plant model $\dot{x}_\mathrm{p} = A_\mathrm{p}x_\mathrm{p} + B_\mathrm{p}u + E_\mathrm{p}w$, $y = C_\mathrm{p}x_\mathrm{p}$,
where $(A_\mathrm{p},B_\mathrm{p},E_\mathrm{p},C_\mathrm{p})$ are real matrices of appropriate dimensions, with $(A_\mathrm{p},B_\mathrm{p})$ stabilisable, and $(A_\mathrm{p},C_\mathrm{p})$ detectable. The plant is stabilised using emulation by the following feedback controller, previously designed in the absence of the wireless network,
$\dot{x}_\mathrm{c} = A_\mathrm{c}x_\mathrm{c} + B_\mathrm{c}y$, $u = C_\mathrm{c}x_\mathrm{c}$,
where $(A_\mathrm{c},B_\mathrm{c},C_\mathrm{c})$ are matrices of appropriate dimensions. We then implement this controller over the network and assume that the in-network processing uses zero-order hold devices so that $\dot{\hat{u}}=0$ and $\dot{\hat{y}}=0$. Also recall that $e=(\hat{y}-y,\hat{u}-u)$ and $x=(x_\mathrm{p},x_\mathrm{c})$. Then, the SHS for the LTI case becomes 
 \eqref{eq:WNCS} with $\mathbf{f}(x,e,w)=A_{11}x + A_{12}e + E_1w$ and $\mathbf{g}(x,e,w)=A_{21}x + A_{22}e + E_2w$, and we denote it by $\Sigma_{\LTI}$,
where $A_{11}\coloneqq  \smallmtx\mtxdd{A_\mathrm{p}}{B_\mathrm{p}C_\mathrm{c}}{B_\mathrm{c}C_\mathrm{p}}{A_\mathrm{c}}$, $A_{12}\coloneqq  \smallmtx\mtxdd{0}{B_\mathrm{p}}{B_\mathrm{c}}{0}$, $E_1\coloneqq  [E_\mathrm{p}^\top\ 0]^\top$,  $A_{21}\coloneqq  - \diag\{C_\mathrm{p},C_\mathrm{c}\}A_{11}$, $A_{22}\coloneqq  -\diag\{C_\mathrm{p},C_\mathrm{c}\}A_{12}$, and $E_2\coloneqq  -\diag\{C_\mathrm{p},C_\mathrm{c}\} E_1$.

The counterpart of Theorem \ref{theo:stability-stoch-protocol} for LTI systems is  below.
\begin{theorem}\label{theo:LTI-stoch}
	Consider the linear WNCS $\Sigma_{\LTI}$ under stochastic scheduling. Then, there exist $\bm{\varepsilon_1,\varepsilon_2,\mu}>0$ and a positive definite symmetric real matrix $\textbf{P}$ such that the LMI in \eqref{eq:LMI-stoch} holds, and an arrival rate $\w\in(0,\infty)$ that satisfies $\w > N\left|\overline{A}_{22}\right|/[\min_{n\in\mathcal{N}} f_n(N-n+1)]$ and
	\begin{align}\label{eq:linear-stochastic}
		\frac{ \sqrt{\bm{\mu}}\big(\sum_{n=1}^{N} \frac{N}{[N-(n-1)]f_n}\big)(1+\rho_{\w}) }{ (\w - \left|\overline{A}_{22}\right|)(1 - \rho_{\w}) } < 1,
	\end{align}
	such that the WNCS \eqref{eq:WNCS} is $\EL_2$ stable in expectation from $w$ to $\left( \overline{A_{21}x},e \right)$ with finite linear gain, where $\rho_{\w}$ is as per \eqref{eq:rho} with $|A|=\left|\overline{A}_{22}\right|$. 
\end{theorem}
\textbf{PROOF:} See Appendix \ref{sec:proof-LTI-stoch}.\qedd

The equivalent of Theorem \ref{theo:Lp-overall} for LTI systems is below.
\begin{theorem}\label{theo:LTI-det}
	Consider the linear WNCS $\Sigma_{\LTI}$ under deterministic scheduling, and suppose the protocol is a.s. UGES with Lyapunov function $W$ and that $|\partial W(k,e)/\partial e|\leq L_1$ for almost all $e\in\R^{n_e}$ and $k\in \N$. Then, there exist $\bm{\varepsilon_3,\varepsilon_4,\vartheta}>0$ and a positive definite symmetric real matrix $\textbf{Q}$ such that the LMI in \eqref{eq:LMI} holds, and an arrival rate $\bm{\omega}$ that satisfies $\w > {L}/{1-\bar{\kappa}}$ and
	\begin{align}\label{eq:numeric-linear}
		\frac{ \sqrt{\bm{\vartheta}} s_{\infty}(\w) }{a_1\w - L_1|A_{22}|} < 1,
	\end{align} 
	such that the linear WNCS $\Sigma_{\LTI}$ is $\EL_2$ stable in expectation from $w$ to $(L_1A_{21}x,W(e))$ with finite linear gain, where $a_1$ is as per \eqref{eq:W-a1a2} and $s_{\infty}$ is as per Theorem \ref{theo:Lp-overall} with $L=L_1|A_{22}|/a_1$.
\end{theorem}
\textbf{PROOF:} The proof follows similar steps to the proof of Theorem \ref{theo:LTI-stoch} and it is thus omitted.\qedd

Theorems \ref{theo:LTI-stoch} and \ref{theo:LTI-det} provide a systematic way of constructing the bound for the transmission rate that ensures stability via the LMI \eqref{eq:LMI-stoch} in case of stochastic scheduling, and the LMI \eqref{eq:LMI} for deterministic scheduling.
We emphasise that for the LTI case, when $w=0$, we can immediately state UGES in expectation (and probability) since \eqref{eq:L1} and \eqref{eq:L2} always hold with $L_1=\max\{|A_{11}|,|A_{12}|\}$ and $L_2=\max\{|A_{21}|,|A_{22}|\}$, see Corollaries \ref{coro:GES} and \ref{coro:GES-in-prob}.
We also note that the LMIs \eqref{eq:LMI-stoch}, \eqref{eq:LMI} could be used to minimise $\bm{\mu},\bm{\vartheta}$ in order to obtain relaxed conditions in \eqref{eq:linear-stochastic} and \eqref{eq:numeric-linear}.

\section{Power control policies}\label{sec:power-control}
We now build upon the stability results from Section \ref{sec:stability} to provide a framework to design transmit power control policies that ensure stability of the non-linear WNCS in Fig. \ref{fig:WNCS}. So far, successful transmission probabilities were treated as given, and now we exploit the fact that these can be modified by adjusting transmit powers.
Recall that throughout this paper, only one node/cluster transmits per transmission instant, and that within a node, there are multiple links that are thus transmitting at the same time and are subject to interference.
Within a node $n$, every link $i_n\in\{1,\dots,\ell_n\}$ has a transmitter and a receiver. 
Let the transmit power of Link $i_n$ be  $p_{n,i_n}\in(0,P_{\max}]$, $P_{\max}\in\R_{> 0}$ is the maximum allowable power at any given time instant, and the channel gain from link's $i_n$ transmitter to link's $j_n$ receiver is given by $g_{n,i_nj_n}>0$. The signal-to-interference-and-noise ratio $\SINR_{n,i_n}$ perceived by the receiver of Link $i_n$ (within Node $n$) is given by \cite{goldsm05}
\begin{align}\label{eq:SINR}
	\SINR_{n,i_n} \coloneqq  \frac{g_{n,i_ni_n}p_{n,i_n}}{\sigma^2_{n,i_n} + \sum_{j_n\neq i_n} g_{n,j_ni_n}p_{n,j_n}},
\end{align}
where $\sigma^2_{n,i_n}\in\R_{> 0}$ is the noise variance at the receiver of Link $i_n$.

The probability of successful reception is a function of $\SINR_{n,i_n}$ (and thus transmit power levels), and we write $f_{n,i_n}$ in Link $i_n$ and Node $n$ as $f_{n,i_n} = \Phi(\SINR_{n,i_n})$ to make such dependence explicit, where $\Phi$ is a non-linear function that represents some model of probability of success. We emphasise that our framework is general and thus does not depend on the choices of the probability model and channel gains. However, we will specify a particular choice of probability model when studying optimal power control further below.

\subsection{Transmit power levels that ensure stability}\label{sec:power-det}
Here, we provide conditions on transmit powers such that the WNCS \eqref{eq:WNCS} is stable for a given average intertransmission time $\bar{\tau}$. We will focus on WNCSs that use deterministic schedules as per Section \ref{sec:protocol-uges}. Particularly, for the sake of illustration, we suppose the RR protocol from Section \ref{sec:protocol-uges} is in place. In this context, a group of sensors/actuators may be scheduled to transmit at any particular timeslot and thus interference may occur, for which we need adequate power control policies.

The lemma below provides a sufficient condition on transmit powers that ensure stability of the WNCS \eqref{eq:WNCS}. It follows from Theorem \ref{theo:Lp-overall} and the proof is thus omitted.
\begin{lemma}
	If the transmit powers $p_{n,1},\dots, p_{n,\ell_n}$, for all $n\in\mathcal{N}$, satisfy
	\begin{align}
		\frac{\gamma \bar{\tau} \textstyle \Big[ \sum_{j=0}^{\infty}   \Big(\frac{1}{1-L\bar{\tau}}\Big)^j\times \prod_{\iota=0}^{j-1}\esp{\kappa_\iota}\Big] }{1 - L\bar{\tau}}<1, \label{eq:power-stability-deter}
	\end{align}
	where $\espp{\kappa_k} = \big( \sum_{n=1}^{N} f_n\psi_n(k) \big)(\eta-1) + 1$, and $f_n =  \prod_{i_n=1}^{\ell_n} \Phi\big(\frac{g_{n,i_ni_n}p_{n,i_n}}{\sigma^2_{n,i_n} + \sum_{j_n\neq i_n} g_{n,j_ni_n}p_{n,j_n}}\big)$.
	Then, the WNCS \eqref{eq:WNCS} under deterministic scheduling is $\EL_p$ stable in expectation for a given average intertransmission time $\bar{\tau}$.\qedd
\end{lemma}

For the single-cluster case $N=1$, we have a closed-form stability bound as per the corollary below.
\begin{corollary}
	For the special case of one node with $\ell$ links, if transmitter powers $p_1,\dots,p_\ell$ satisfy
	\begin{align}
		\prod_{i=1}^{\ell}\Phi\bigg( \frac{g_{ii}p_{i}}{\sigma^2_{i} + \sum_{j\neq i} g_{ji}p_{j}} \bigg) > \frac{\bar{\tau}(\gamma + L)}{1- \eta}, \label{eq:constraint-det-1node}
	\end{align}
	then, the WNCS \eqref{eq:WNCS} is $\EL_p$ stable in expectation for a given average intertransmission time $\bar{\tau}$.\qedd
\end{corollary}


\subsection{Power control as an optimization problem}\label{sec:powers-opt}
The above results give conditions on the transmit power levels to ensure stability. In practice, we are also interested in minimising these powers, and we explain how to do it here by formulating corresponding optimisation problems.

\begin{problem}[$N$ nodes, $\ell_n$ links per node]\label{prob:optimal-powers}
	Let $\mathbf{p}\coloneqq (p_{1,1},\dots,p_{1,\ell_1},\dots,p_{N,1},\dots,p_{N,\ell_N})\in\R^{\ell_1+\cdots+\ell_N}$ be the vector containing all transmit powers.
	Consider the WNCS \eqref{eq:WNCS}, and the weights $\mathfrak{w}_{n,i_n} \in \R_{\geq 0}$, for all $n\in\mathcal{N}$ and $i_n\in\mathcal{C}_n\coloneqq \{1,\dots,\ell_n\}$. Find transmit powers $\mathbf{p}$ by solving the  optimisation problem $\min_{\mathbf{p}\in\R^{\ell_1+\cdots +\ell_N}}\quad  \sum_{n\in\mathcal{N}} \sum_{i_n\in\mathcal{C}_n} \mathfrak{w}_{n,i_n} p_{n,i_n}$
	subject to $p_{n,i_n} \in (0,P_{\max}]$, for all $n\in\mathcal{N}$ and $i_n\in\mathcal{C}_n$, and the stability constraint \eqref{eq:power-stability-deter}.\fin
\end{problem}
Problem \ref{prob:optimal-powers} is hard to solve analytically, but one can attempt to solve it numerically, although the stability constraint \eqref{eq:power-stability-deter} is highly non-linear.
A simpler optimisation problem can be derived from Problem 1 in the special case of finding transmitter powers for the links in one node only. To that end, we write the stability condition \eqref{eq:constraint-det-1node} as non-strict inequalities. This will slightly introduce conservatism in the optimal solution with respect to an alternative optimisation problem that uses the strict stability inequality; however, we will show that it comes with a great advantage in terms of computational load.

\begin{problem}[One node, $\ell$ links]\label{prob:1-node-L-channels}
	Consider the WNCS \eqref{eq:WNCS} with one node ($N=1$) and $\ell$ links, find transmit powers $\mathbf{p}=(p_1,\dots,p_\ell)$ by solving the optimisation problem $\min_{\mathbf{p}\in\R^{\ell}}\ \sum_{i=1}^{\ell} p_i$
	subject to $p_i \in (0,P_{\max}], \forall i\in\{1,\dots,\ell\}$, and the stability constraint
	\begin{align}\label{eq:stab-constraint}
		\textstyle 
		\prod_{i=1}^{\ell}\Phi(\text{SINR}_i)\geq 
		\frac{\bar{\tau}(\gamma + L)}{1- \eta}  -\delta,
	\end{align}
	for some sufficiently small $\delta>0$.\fin
\end{problem}
Problem \ref{prob:1-node-L-channels} is easier to solve than Problem 1, and we will show next that it can be solved via a family of linear programs for a specific choice of probability model $\Phi$. We choose a model often used in information-theoretic outage analysis \cite{biglieri1998fading}, which is defined as $\Phi(\text{SINR}_i)\coloneqq \exp(-a/\text{SINR}_i)$, $i\in\{1,\dots,\ell\}$, where $a>0$. 
Let us first introduce some preliminary definitions. $\mathscr{C}_{\bar{\tau}}\coloneqq \ln\big( \big[ \tfrac{\bar{\tau}(\gamma + L)}{1- \eta}  -\delta \big]^{-1/a} \big)$, $\mathbf{1}\coloneqq (1,\dots,1)\in\R^\ell$, and $\mathscr{Q}\coloneqq \{q\in\R^{\ell}:q_i\in(0,1)\wedge \sum_{i=1}^{\ell} q_i=1\}$.

\begin{proposition}\label{propo:equivalent-LP}
	For the above choice of probability model, the solution to Problem \ref{prob:1-node-L-channels}  can be computed as follows.
	\begin{enumerate}
		\item Given any $q\in\mathscr{Q}$, solve the below linear program (provided it is feasible),
		\begin{subequations}\label{eq:LP}
			\begin{align}
				&\min_{\mathbf{p}\geq 0}\ \mathbf{1}^\top \mathbf{p}  \\
				&\text{s.t.}\ \mathbf{A}(q)\mathbf{p} \geq \mathbf{b}, \label{eq:LP-constraint}
			\end{align}
		\end{subequations}
		where
		\begin{align*}
			\mathbf{A}(q)\coloneqq \smallmtx \begin{bmatrix}
				g_{11}q_1\mathscr{C}_{\bar{\tau}} & \cdots & -g_{\ell 1} \\
				\vdots & \ddots & \vdots \\
				-g_{1\ell} & \cdots & g_{\ell\ell}q_\ell \mathscr{C}_{\bar{\tau}} \\
				-1 & \cdots & 0 \\
				\vdots & \ddots & \vdots \\
				0 & \cdots & -1
			\end{bmatrix},
			\mathbf{b}\coloneqq \begin{bmatrix}
				\sigma_1^2 \\ \vdots \\ \sigma_\ell^2\\ -P_{\max} \\ \vdots\\ -P_{\max}
			\end{bmatrix}.
		\end{align*}
		\item Let $\mathbf{p}^*({q})$ denote the solution to \eqref{eq:LP} for any given $q$. Solve $q^* = \arg\min_{q\in\mathscr{Q}}\mathbf{p}^*(q)$. Then, the solution to Problem 2 is given by $\mathbf{p}^*=\mathbf{p}^*(q^*)$.
	\end{enumerate}
\end{proposition}
\textbf{PROOF:} See Appendix \ref{sec:proof-equiv-LP}.\qedd

Proposition \ref{propo:equivalent-LP} states that we can solve Problem \ref{prob:1-node-L-channels} by solving a family of linear programs parametrised in $q$. Since LPs are computationally inexpensive, we can create a fine grid for $q\in\mathscr{Q}$ and solve an LP \eqref{eq:LP} for each one of these grid points and check which one has the minimum cost function (cf. Part 2 of Proposition \ref{propo:equivalent-LP}). Note that feasibility of each LP depends on the average inter-transmission time $\bar{\tau}$, maximum available power $P_{\max}$, and channel interference via the channel gains $g_{ij}$, which foreshadows a trade-off between communication constraints and stability when designing optimal power policies. For the two-link case, i.e. $\ell=2$, the LP formulation allows us to state explicit feasibility conditions and thus solve Problem \ref{prob:1-node-L-channels} in closed-form, see Proposition \ref{propo:optimisation} and Corollary \ref{coro:epsilon} below.

In the wireless telecommunications community, LPs have appeared in the literature when considering power control problems that do not consider closed-loop stability. Instead, these results are interested in optimising QoS and network capacity. For instance, in \cite[Chapter 2]{chiang2008power}, the problem of optimal distributed power control subject to fixed target SINR constraints can be written as an LP, where the constraints are individual SINR requirements. Here, with the aim of bridging the gap between telecommunications and control, we showed that the problem of using minimal transmission power to achieve a control-oriented objective such as stability can also be written as an LP, and thus standard techniques such as the ones presented in \cite{chiang2008power} can be used to solve them.

The two-link results are presented below.
\begin{proposition}\label{propo:optimisation}
	Consider the linear program \eqref{eq:LP} for $\ell=2$, $q_1=\varepsilon$, and $q_2=1-\varepsilon$, with
	\begin{align}\label{eq:interval}
		\textstyle 
		\varepsilon\in\Big(\frac{1}{2}-\Big[\frac{1}{4}-\frac{g_{12}g_{21}}{g_{11}g_{22}\mathscr{C}_{\bar{\tau}}^2}\Big]^{1/2},\frac{1}{2}+\Big[\frac{1}{4}-\frac{g_{12}g_{21}}{g_{11}g_{22}\mathscr{C}_{\bar{\tau}}^2}\Big]^{1/2}\Big).
	\end{align}
	If $\bar{\tau}< \frac{1-\eta}{\gamma + L}[\delta + \exp(-a\sqrt{4g_{12}g_{21}/(g_{11}g_{22})})]$ 
	and $P_{\max} \geq \max\{p_1^*(\varepsilon),p_2^*(\varepsilon)\}$, where
	\begin{subequations}\label{eq:powers-epsilon}
		\begin{align}
			p_1^*(\varepsilon) & = \frac{-\varepsilon g_{22}\mathscr{C}_{\bar{\tau}}\sigma_1^2 +g_{21}\sigma_2^2 + g_{22}\mathscr{C}_{\bar{\tau}}\sigma_1^2 }{-\varepsilon^2g_{11}g_{22}\mathscr{C}_{\bar{\tau}}^2 + \varepsilon g_{11}g_{22}\mathscr{C}_{\bar{\tau}}^2 - g_{12}g_{21}}, \\
			p_2^*(\varepsilon) &= \frac{\varepsilon g_{11}\mathscr{C}_{\bar{\tau}}\sigma_2^2+g_{12}\sigma_1^2}{-\varepsilon^2g_{11}g_{22}\mathscr{C}_{\bar{\tau}}^2 + \varepsilon g_{11}g_{22}\mathscr{C}_{\bar{\tau}}^2 - g_{12}g_{21}}.
		\end{align}
	\end{subequations}
	Then, \eqref{eq:powers-epsilon} solves the LP \eqref{eq:LP} for $\ell=2$.
\end{proposition}
\textbf{PROOF:} See Appendix \ref{sec:proof-two-channel}.\qedd

The feasibility conditions for the two-link case translate into bounds on the average inter-transmission time $\bar{\tau}$ and the maximum available power $P_{\max}$. This is consistent since we should not be able to transmit as slowly as we want (large $\bar{\tau}$) and expect to have a stable WNCS under minimum power. 

Proposition \ref{propo:optimisation} provides the parametrised optimal powers $\mathbf{p}(\varepsilon)$ that solve the LP \eqref{eq:LP} for $\ell=2$. In the next corollary, we  find the optimal $\varepsilon^*$  for which $\mathbf{p}^*=\mathbf{p}^*(\varepsilon^*)$ solves the original Problem \ref{prob:1-node-L-channels} with $\ell=2$.

\begin{corollary}\label{coro:epsilon}
	Let $\varepsilon^*=\arg\min_{\varepsilon}p_1^*(\varepsilon)+p_2^*(\varepsilon)$, where the search is done in the interval \eqref{eq:interval}. Then, $\varepsilon^*$ can be computed as the solution to $\mathfrak{a}\varepsilon^2+\mathfrak{b}\varepsilon+\mathfrak{c}=0$ in the interval \eqref{eq:interval}, with
	\begin{align*}
		\mathfrak{a} &\coloneqq g_{11}^2g_{22}\mathscr{C}_{\bar{\tau}}^3\sigma_2^2 - g_{11}g_{22}^2\mathscr{C}_{\bar{\tau}}^3\sigma_1^2, \\
		\mathfrak{b} &\coloneqq 2g_{11}g_{22}^2\mathscr{C}_{\bar{\tau}}^3\sigma_1^2 + 2g_{11}g_{12}g_{22}\mathscr{C}_{\bar{\tau}}^2\sigma_1^2 + 2g_{11}g_{21}g_{22}\mathscr{C}_{\bar{\tau}}^2\sigma_2^2, \\
		\mathfrak{c} &\coloneqq g_{12}g_{21}g_{22}\mathscr{C}_{\bar{\tau}}\sigma_1^2  -g_{11}g_{22}^2\mathscr{C}_{\bar{\tau}}^3\sigma_1^2 - g_{11}g_{12}g_{22}\mathscr{C}_{\bar{\tau}}^2\sigma_1^2 \\
		&\qquad - g_{11}g_{21}g_{22}\mathscr{C}_{\bar{\tau}}^2\sigma_2^2 - g_{11}g_{12}g_{21}\mathscr{C}_{\bar{\tau}}\sigma_2^2.
	\end{align*}
	Moreover, $\mathbf{p}^*=(p_1^*(\varepsilon^*),p_2^*(\varepsilon^*))$ are the optimal powers that solve Problem \ref{prob:1-node-L-channels}.
\end{corollary}
\textbf{PROOF:} The proof follows immediately from minimising $p_1^*(\varepsilon)+p_2^*(\varepsilon)$ in the interval \eqref{eq:interval}, where $p_1^*(\varepsilon)$ and $p_2^*(\varepsilon)$ are as per \eqref{eq:powers-epsilon}. Particularly, $\mathfrak{a}\varepsilon^2+\mathfrak{b}\varepsilon+\mathfrak{c}=0$ follows from  $d(p_1^*(\varepsilon)+p_2^*(\varepsilon))/d\varepsilon=0$.\qedd

For the case with equal noises $\sigma_1^2=\sigma_2^2=\sigma^2$ and equal channel gains $g_{11}=g_{12}=g_{21}=g_{22}=g$, it is not hard to see that $\varepsilon$ in Corollary \ref{coro:epsilon} is equal to $\varepsilon= 1/2$, and thus the optimal powers are given by $p_1^*=p_2^*=\frac{2\sigma^2}{(\mathscr{C}_{\bar{\tau}}-2)g}$ provided $\bar{\tau}< \frac{1-\eta}{\gamma + L}[\delta + \exp(-2a)]$ and  $P_{\max}\geq \frac{2\sigma^2}{(\mathscr{C}_{\bar{\tau}}-2)g}$.

\section{Numerical examples}\label{sec:numerical}
We illustrate our results in the control of an unstable batch reactor over a wireless network. The batch reactor is a two-input two-output system of the form $\dot{x}_\mathrm{p} = A_\mathrm{p}x_\mathrm{p} + B_\mathrm{p}u$, $y = C_\mathrm{p}x_\mathrm{p}$, with  $(A_\mathrm{p},B_\mathrm{p},C_\mathrm{p})$ given in \cite[p.62]{green12}. We design a standard observer-based feedback controller that stabilises the batch reactor in the absence of the network. This leads to $\dot{x}_\mathrm{c} = A_\mathrm{c}x_\mathrm{c} + B_\mathrm{c}y$, $u = C_\mathrm{c}x_\mathrm{c}$, with
\begin{align*}
	A_\mathrm{c} &= \smallmtx \begin{bmatrix}
		-7.88 &  -4.86 &  -2.55 &   3.59\\
		-0.72 &  -6.60 &  -0.59 &   0.65\\
		-11.61&  -23.35&  -13.19&   11.92\\
		-10.43&  -21.57&   -9.22&    8.34
	\end{bmatrix}, \\
	B_\mathrm{c} &= \smallmtx	\begin{bmatrix}
		9.26 & 0.85 & 10.97 & 10.62\\
		4.65 & 4.72 & 27.28 & 26.33
	\end{bmatrix}^\top , \\
	C_\mathrm{c} &= \smallmtx \begin{bmatrix}
		0.13  &  0.42 &    0.046 &  -0.15\\
		0.59  &  0.26 &  -1.39  &  1.52
	\end{bmatrix}.
\end{align*}
The controller is then implemented over the network and the resulting WNCS model has the form $\Sigma_{\LTI}$ with $E_1=0$, $E_2=0$ (cf. Section \ref{sec:linear}).
Since we do not consider disturbances in this example, the stability notion used throughout Section \ref{sec:numerical} will implicitly be UGES in expectation (or in probability, see Corollaries \ref{coro:GES} and \ref{coro:GES-in-prob}).

%
\subsection{Transmission bounds for given success probabilities}
Here we compute the arrival rate that ensures stability for a given wireless network with fixed probabilities of success.
Consider two nodes $N=2$, and two links $\ell_1=\ell_2=2$ per node.  
We send the two output measurements through the two links in Node 1, and the two control inputs through the two links in Node 2. Therefore, $e=(e_1,e_2)$, where $e_1=(\hat{y}_1-y_1,\hat{y}_2-y_2)$ and $e_2=(\hat{u}_1-u_1,\hat{u}_2-u_2)$.
Consider the probabilities of success $f_{1,1}=0.3, f_{1,2}=0.8,f_{2,1}=0.75$, and $f_{2,2}=0.8$. These imply that the cumulative success probabilities of nodes one and two are $f_1=0.3\times 0.8 =0.24 $ and $f_2=0.75\times 0.8=0.6$, respectively.

\textbf{Stochastic scheduling:}
We use Theorem \ref{theo:LTI-stoch} to find the transmission rate that ensures stability under stochastic scheduling. The required parameters are as follows. We solve the LMI \eqref{eq:LMI-stoch} with \textsc{Yalmip} and obtain $\bm{\mu}=472.68$. Also, $|\overline{A}_{22}|=8.8$.
Then, from Theorem \ref{theo:LTI-stoch}, the transmission bound $\w_{\text{\tiny Thm.\ref{theo:LTI-stoch}}}$ that ensures stability is given by $\w_{\text{\tiny Thm.\ref{theo:LTI-stoch}}}=187.29$. Recall that $1/\w_{\text{\tiny Thm.\ref{theo:LTI-stoch}}}$ represents the average intertransmission time that ensures stability. That is, stability is guaranteed provided transmissions occur every $1/187.29=5.3$[ms] in average. We compare our bound to the previous work \cite{tabnes08-tac}, which ensures stability in our setting with a larger transmission rate, see Lemma \ref{lem:tabbara}. This rate is given by $\w^{\text{\tiny stoch.}}_{\text{ \cite{tabnes08-tac}}}=305.11$, which is 1.63 times more conservative than our bound that exploits the structure of the wireless network through multiple links with possibly different success probabilities.

\textbf{Deterministic scheduling:}
We use Theorem \ref{theo:LTI-det} for a round-robin deterministic scheduling. The required parameters are as follows.
We solve the LMI \eqref{eq:LMI} and obtain $\bm{\vartheta}=940.96$. From Lemma \ref{lem:RR} we have $a_1=1$, $\eta=\sqrt{0.5}$ and $\bar{\kappa}=0.93$, and $\esp{\kappa_k} = 0.93$ for $k = 1 + 2\sigma$, and $\esp{\kappa_k} = 0.82$ for $k = 2 + 2\sigma$, $\sigma\in\N$. Similar to \cite{nestee04}, it is not hard to find $L_1=\sqrt{N}=1.41$. Also, $|A_{22}|=8.8$. 
Then, using Theorem \ref{theo:LTI-det} we compute the arrival rate $\w_{\text{\tiny Thm.\ref{theo:LTI-det}}}=275.31$, which translates into an average transmission of $3.6$[ms]. Next, we compute the bound given in \cite{tabnes08-tac}, i.e. $\w^{\text{\tiny deter.}}_{\text{ \cite{tabnes08-tac}}}=613.34$, which is 2.23 times more conservative than our bound $\w_{\text{\tiny Thm.\ref{theo:LTI-det}}}$.

We can see that for both stochastic and deterministic protocols, our bounds are around twice less conservative than the previous available---and more restrictive ones---in \cite{tabnes08-tac}, which highlights our results (cf. Lemma \ref{lem:tabbara}).

\subsection{Transmit power levels}\label{sec:num-power}
We now illustrate the results on power control from Section \ref{sec:power-control}. We first focus on applying Proposition \ref{propo:optimisation} and Corollary \ref{coro:epsilon} to compute optimal powers. To that end, we consider the batch reactor sends the measurements to the controller via two links, and the control inputs are directly available to the plant (this fits Problem \ref{prob:1-node-L-channels} since we have one node and two links). 
The matrices $A_{12},A_{21},A_{22}$ in the WNCS model $\Sigma_{\LTI}$ change slightly for this case, since now $e= (\hat{y}_1-y_1,\hat{y}_2-y_2)$, and are given by $A_{12}=[0\ B_\mathrm{c}^\top]^\top$, $A_{21}=-[C_\mathrm{p}\ 0]A_{11}$, and $A_{22}=-[C_\mathrm{p}\ 0]A_{12}$. 
Consider the channel gains $g_{11}=0.2,g_{22}=0.063$, and $g_{12}=g_{21}=0.012$. We pick $\sigma_1^2=\sigma_2^2=1$, $P_{\max}=70$ [W], and $a=1$. 
The average intertransmission time $\bar{\tau}$ should be picked as per Proposition \ref{propo:optimisation}, i.e. $\bar{\tau}<0.0205$. We choose $\bar{\tau}=5$[ms], which leads to $\mathscr{C}_{\bar{\tau}}=1.62$. We now use Corollary \ref{coro:epsilon} to find $\varepsilon^* = 0.38$. Therefore, via Proposition \ref{propo:optimisation}, the optimal powers are given by $p_1^*(\varepsilon^*) = 9.9$ and $p_2(\varepsilon^*)=17.6$, which are feasible since $P_{\max}=70$. 

\begin{figure}
	\centering
	\includegraphics[scale=0.7]{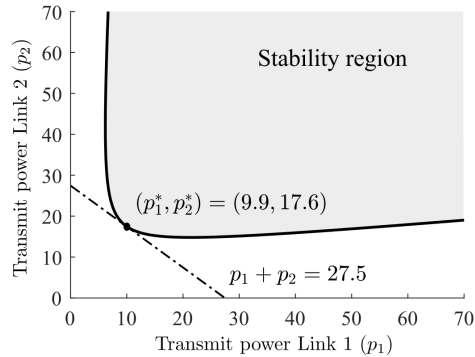}
	\caption{Stability region and the optimal power policy.}
	\label{fig:region}
\end{figure}

For illustration purposes, we show the stability region \eqref{eq:stab-constraint} in Fig. \ref{fig:region}. Any transmitter powers inside that region will ensure stability of the WNCS. The results in Section \ref{sec:power-det} provided a methodology to numerically construct this region. Then, in Section \ref{sec:powers-opt} we provided an optimisation framework to ensure stability with minimum power, which led to the data-tip in Fig. \ref{fig:region} with the computations shown above.

Our results illustrate that more advanced strategies to design transmit power levels are important when compared to naive strategies such as implementing maximum power. Using maximum power may not be viable since increasing power in turn increases the interference between links, which may worsen the probability of success. In addition, maximum power causes serious energy consumption and electromagnetic pollution. For instance, using our optimal strategy $(p_1,p_2)=(9.9,17.6)$ leads to about $86\%$ less power usage at transmitter one and $75\%$ at transmitter two, when compared to a strategy that uses $(p_1,p_2)=(P_{\max},P_{\max})=(70,70)$.

To finalise, we show how to use the LP approach in Proposition \ref{propo:equivalent-LP}. To that end, we treat the two sensors and two actuators as one node with four wireless links, i.e. $\ell=4$ in Proposition \ref{propo:equivalent-LP}.  Consider the channel gains $g_{11}=g_{33}=0.2$, $g_{22}=0.3$, $g_{44}=0.4$, and $g_{12}=g_{21}=0.063,g_{13}=g_{31}=0.012,g_{14}=g_{41}=0.05$. We pick $\bar{\tau}=1$[ms]. Then, as per Proposition \ref{propo:equivalent-LP}, the optimal powers are $\mathbf{p}^*=(15,10,14,11)$, which again, are a significantly better alternative to maximum power usage.

\section{Conclusions}\label{sec:conclusion}
We provided a general framework for the study of stability and power control for non-linear WNCS with multiple wireless links subject to stochastic packet losses and random transmission instants. The results from this paper contribute to closing the gap between control and communication literature by carefully modelling the wireless network and connecting the stability results to transmit power control via realistic and well-known interference models.
Different packet loss models such as Markovian losses are also of interest for future work.

\appendix
%
\section{Proof of Theorem \ref{theo:stability-stoch-protocol}}\label{sec:proof-stab-stoch-prot}
The following preliminary results are needed to prove Theorem \ref{theo:stability-stoch-protocol}. 

\begin{lemma}\label{lem:QeAt}
	Suppose $\tau_k$ satisfies Assumption \ref{assu:transmission-instants}. Let $T$ be the cover time for the sequence $\{(Q(0),\tau_0),\dots,(Q(T-1),\tau_{T-1})\}$. Then, the following holds.
	\begin{enumerate}
		\item $\big|\prod_{k=0}^{T-1} Q(k)\exp(A\tau_k)\big| \leq \exp\big( |A|\sum_{k=0}^{T-1} \tau_k \big)-1$.
		\item $\esp{Z} = \prod_{n=1}^{N} \frac{ (N-(n-1))f_n }{ N(1-|A|/\w-(1-f_n)) - (n-1)f_n }$, with $\w > N|A|/[\min_{n\in\mathcal{N}} f_n(N-n+1)]$, and $Z\coloneqq  \exp\big( |A|\sum_{k=0}^{T-1} \tau_k \big)$.
		\item $\exists \w\in(0,\infty)$ s.t. $\textstyle\espp{\big|\prod_{k=0}^{T-1} Q(k)\exp(A\tau_k)\big|} < 1$.
	\end{enumerate}
\end{lemma}
\textbf{PROOF:}\\
(1) Follows directly from \cite[Lemma 9.1]{tabnes08-tac}.\\
(2) Let $\tilde{\tau} = \sum_{k=0}^{T-1} \tau_k$. Via Example 1.8.13 in \cite{stir05}, we can compute the moment generating function of $S$ as $M_{\tilde{\tau}}(s) = \espp{\exp(s\tilde{\tau})} = G_T(M_{\tau}(s))$,
	where $M_\tau$ is the moment generating function of the exponentially distributed random variables $\tau_k$. Particularly, $M_\tau(s) = \w/(\w-s)$, for $\w>s$. Therefore,
	$\espp{Z} = \espp{\exp\left( |A|\tilde{\tau} \right)} = 	G_T(\w/(\w-|A|))$, $\w>|A|$,
	where $G_T(s)$ is given in Lemma \ref{lem:pgf-T}. Note that we require both $\w/(\w-|A|) < \frac{1}{1-[\min_{n\in\mathcal{N}} f_n(N-n+1)]/N}$ and $\w>|A|$. The proof is complete by noting that choosing $\w$ as per the theorem statement satisfies both bounds.
	\\
(3) From Parts 1) and 2), and for $\w\to\infty$, we get $\lim_{\w\to\infty} \textstyle \esp{|\prod_{k=0}^{T-1} Q(k)\exp(A\tau_k)|} \leq \lim_{\w\to\infty} \espp{Z} - 1 = 0$.
	%
	%
	Define $\mathfrak{f}(\w) \coloneqq  \esp{|\prod_{k=0}^{T-1} Q(k)\exp(A\tau_k)|}>0$.
	Note that $\esp{Z}-1$ is a strictly decreasing function of $\w$ for $\w>0$, which upper bounds $\mathfrak{f}(\w)$. Then, $\lim_{\w\to\infty} \mathfrak{f}(\w) = 0$ implies that $\exists \w^*\in(0,\infty)$ such that for all $\w>\w^*$, $\mathfrak{f}(\w) <1$, concluding the proof.
	\qedd

To conclude the proof we use a small-gain argument. That is, we look at the WNCS \eqref{eq:WNCS} as the interconnection of two subsystems, namely the $x$-subsystem and the $e$-subsystem, and we state a small-gain condition that will ensure overall stability. First we state $\EL_p$ stability in expectation for the $e$-subsystem via the proposition below.
\begin{proposition}\label{propo:Lp-e-stochastic}
	Consider the WNCS \eqref{eq:WNCS} under a stochastic protocol (a.s. covering), and suppose the following holds.
	\begin{enumerate}[$(i)$]
		\item Item $(i)$ of Theorem \ref{theo:stability-stoch-protocol} is satisfied.
		\item $\w$ is chosen such that $\rho_{\w} < 1$, where $\rho_{\w}\coloneqq  \espp{Z}-1$, and $\espp{Z}$ is as per Lemma \ref{lem:QeAt}.
	\end{enumerate}
	Then, the $e$-subsystem \eqref{eq:e-sys},\eqref{eq:protocol} is $\EL_p$ stable in expectation from $\tilde{y}$ to $e$ with gain $\gamma_{\w}=\frac{\mathbf{E}\{T\}(1+\rho_{\w})}{(\w - |A|)(1-\rho_{\w})}$, where $\espp{T}$ is as per Lemma \ref{lem:pgf-T}.
\end{proposition}
\textbf{PROOF:} Having computing all the necessary preliminaries in Lemma \ref{lem:QeAt}, the proof of Proposition \ref{propo:Lp-e-stochastic} follows exactly as the proof of Theorem 9.4 in \cite{tabnes08-tac}, and it is thus omitted given space constraints.\qedd

Since Proposition \ref{propo:Lp-e-stochastic} shows the $e$-subsystem is $\EL_p$ stable in expectation from $\tilde{y}(x,w)$ to $e$, and from the statement of Theorem \ref{theo:stability-stoch-protocol} the $x$-subsystem is $\EL_p$ stable in expectation from $(e,w)$ to $G(x)$, we can appeal to the small-gain theorem in e.g. \cite[Theorem 1]{nestee04} to state that $\gamma \gamma_{\w}< 1$ ensures $\EL_p$ stability of \eqref{eq:WNCS}, concluding the proof.\qedd

\section{Proof of Theorem \ref{theo:Lp-overall}}\label{sec:appendix-stability-det}
Similarly to the proof of Theorem \ref{theo:stability-stoch-protocol}, to state the stability result, we will use a small-gain approach. That is, we first present a proposition that provides
sufficient conditions for $\mathcal{L}_p$ in expectation of the error subsystem \eqref{eq:e-sys}--\eqref{eq:protocol}, and then use this result together with small-gain arguments to show $\EL_p$ in expectation of the overall system \eqref{eq:WNCS}.

\begin{proposition}\label{propo:Lp-e-sys}
	Consider the WNCS \eqref{eq:WNCS} and suppose the following holds.
	\begin{enumerate}[$(i)$]
		\item System \eqref{eq:protocol} is a.s. UGES with Lyapunov function $W$.
		\item There exists $L\geq 0$ such that
		\begin{align}\label{eq:dWdt-propo}
			\left\langle {\partial W}/{\partial e},\mathbf{g}(x,e,w)\right\rangle \leq LW(k,e) + |\tilde{y}(x,w)|
		\end{align}
		holds for almost all $e\in\R^{n_e}$, all $(x,w)\in\R^{n_x}\times\R^{n_w}$, $t\in(t_k,t_{k+1})$, and $k\in\N$, where $\tilde{y}:\R^{n_x}\times\R^{n_w}\rightarrow\R^{n_e}$ is a continuous function of $(x,w)$.
	\end{enumerate}
	If $\w > {L}/{1-\bar{\kappa}}$,
	then the error subsystem \eqref{eq:e-sys}, \eqref{eq:protocol} is $\mathcal{L}_p$ stable in expectation from $\tilde{y}(x,w)$ to $W(e)$ with finite  gain $\gamma_e \coloneqq  {s_{\infty}(\w)}/{(\w-L)}$.
\end{proposition}
\textbf{PROOF:} Similarly to \cite[Section X]{nestee04}, inequality \eqref{eq:dWdt-propo} implies
\begin{align}\label{eq:dWdt-proof}
	dW(k,e(t))/dt \leq LW(k,e(t)) +  |\tilde{y}|,
\end{align}
for almost all $t$ and for all $k\in\N$. Set $t_{-1}=0$. We can proceed similar to the proof of \cite[Proposition 9.5]{tabnes08-tac} by applying standard comparison lemmas to the above inequality, using \eqref{eq:W-kappa}, and iterating the resulting linear recurrence to obtain
\begin{multline}\label{eq:W+-recur-proof}
	W(k+1,e(t_k^+)) \leq  \bigg( \prod_{j=0}^{k} R_j \bigg) W(0,e(0)) \\ + \exp(-L\tau_k)\sum_{j=0}^{k} \bigg(\prod_{i=j}^{k} R_i\bigg)\int_{t_{j-1}}^{t_j} \exp(L(t_j-s))|\tilde{y}(s)|\mathrm{d}s
\end{multline}
for all $k\in\N$,  where $R_k\coloneqq  \kappa_k \exp(L\tau_k)$. Similar to \cite[Proposition 9.5]{tabnes08-tac}, we set $\tilde{y}=0$, and we can obtain the following inequality for all $M\in\N$ and $p\in\N\cup\{+\infty\}$
\begin{align}\label{eq:Wp-0}
	\norm{W}_{\EL_p[t_M,t_{M+1}]} \leq \textstyle \Big(\prod_{j=0}^{M}R_j \Big)\frac{\exp(L\tau_{M+1})}{\min\{L,1\}}W(0,e(0)).
\end{align}
Note that,
\begin{align}
	\textstyle 
	\espp{\prod_{j=0}^{M}R_j} {=} \textstyle  \left( \frac{\w}{\w-L} \right)^{M+1} \prod_{j=0}^{M}\espp{\kappa_j}.
	\label{eq:Esp-Rj}
\end{align}
Therefore, applying the expectation operator into \eqref{eq:Wp-0} and using \eqref{eq:Esp-Rj} we get $\espp{\norm{W}_{\EL_p[t_M,t_{M+1}]}} \leq \textstyle \big( \frac{\w}{\w-L} \big)^{M+2}\big(\prod_{j=0}^{M}\espp{\kappa_j}\big)  \big(\frac{W(0,e(0))}{\min\{L,1\}}\big)$.
From the above inequality, and by using linearity of the expectation operator, we have that $\espp{\norm{W}_{\EL_p[0,t]}} \leq \textstyle  \big(\frac{W(0,e(0))}{\min\{L,1\}}\big) \big(\frac{\w}{\w-L}\big) \sum_{j=0}^{\infty}\big[ \big(\frac{\w}{\w-L}\big)^j \times \prod_{k=0}^{j-1} \espp{\kappa_k}\big]$.
Now, since the underlying protocol is a.s. UGES, we have that $\esp{\kappa_j}\leq \bar{\kappa}$ for all $j\in\N$. We can thus bound the above as
\begin{align}\label{eq:W0-contribution}
	\espp{\norm{W}_{\EL_p[0,t]}} &\leq \textstyle \left(\frac{W(0,e(0))}{\min\{L,1\}}\right)\left( \frac{\w}{\w(1-\bar{\kappa})-L} \right),
\end{align}
where we used the fact that the corresponding geometric series converges since $\w > {L}/{1-\bar{\kappa}}$ holds.

We now move to the contribution of the disturbance term $\tilde{y}$, and we hence set $W(0,e(0))=0$. From \eqref{eq:W+-recur-proof},
\begin{multline}\label{eq:W+-proof}
	W(k+1,e(t_k^+)) \leq \textstyle  \exp(-L\tau_k)\sum_{j=0}^{k} \Big(\prod_{i=j}^{k} R_i\Big) \\
	\times \int_{t_{j-1}}^{t_j} \exp(L(t_j-s))|\tilde{y}(s)|\mathrm{d}s.
\end{multline}
Applying variations of parameters formula (see e.g. Lemma 13 in \cite{manepo19}) to \eqref{eq:dWdt-proof} with initial condition \eqref{eq:W+-proof}, 
\begin{multline}\label{eq:W-variations}
	W(k+1,e(\theta)) \leq \exp(-L\tau_k)\exp(L(\theta-t_k)) \\
	 \times\sum_{j=0}^{k} \bigg(\prod_{i=j}^{k} R_i\bigg) \int_{t_{j-1}}^{t_j} \exp(L(t_j-s))|\tilde{y}(s)|\mathrm{d}s  \\
	 + \int_{t_k}^{\theta} \exp(L(\theta - s ))|\tilde{y}(s)|\mathrm{d}s,
\end{multline}
for all $\theta\in[t_k,t_{k+1}]$. We take expectation of the supremum of the bound in \eqref{eq:W-variations}, which yields the following
\begin{multline}\label{eq:Linf-bound}
	\espp{\norm{W}_{\EL_{\infty}[t_k,t_{k+1}]}} 
	%
	\leq \textstyle  \espp{\norm{\varphi}_{\EL_1[0,\tau]}}\sum_{j=0}^{k+1}\Big[ \\
	\textstyle \left(\frac{\w}{\w-L}\right)^{k-j+1}\Big( \prod_{\iota=0}^{k-j} \espp{\kappa_{\iota}}\Big)\espp{\norm{\tilde{y}}_{\EL_{\infty}[t_{j-1},t_j]}} \Big] ,
\end{multline}
where we also used \eqref{eq:Esp-Rj}, the fact that $\tau_k$ is stationary thus we dropped the subscript, and $\varphi(s)\coloneqq  \exp(Ls)$.
We use $\exp(L(\theta-t_k))\leq \exp(L\tau_{k})$
in \eqref{eq:W-variations} to get \sloppy 
$\espp{\norm{W}_{\EL_1[t_k,t_{k+1}]}} 
	\leq \textstyle  \espp{\norm{\varphi}_{\EL_1[0,\tau]}}\sum_{j=0}^{k+1}\Big[\left(\frac{\w}{\w-L}\right)^{k-j+1} \prod_{\iota=0}^{k-j} \espp{\kappa_{\iota}}\times\espp{\norm{\tilde{y}}_{\EL_{1}[t_{j-1},t_j]}} \Big]$,
where we have used the Young's inequality and the H\"older's inequality, as in \cite{tabnes08-tac}, to split the integrals. Note that $\espp{\norm{\varphi}_{\EL_1[0,\tau]}} = {\espp{\exp(L\tau)}}/{L} - {1}/{L}= {1}/{\w - L}$.
Now, similar to \cite{tanete07}, since the $\EL_1$ and $\EL_{\infty}$ norms are bounded by the same expressions, we can use the Riesz-Thorin theorem (see \cite[Theorem 1.2]{tanete07}) to bound,
for all $p\in\N\cup\{+\infty\}$, 
\begin{align*}
	\espp{\norm{W}_{\EL_p[t_k,t_{k+1}]}} &\leq \textstyle  \frac{1}{\w-L}\sum_{j=0}^{k+1}\Big[\left(\frac{\w}{\w-L}\right)^{k-j+1}\non \\
	&\hspace{-1cm}\times \textstyle  \prod_{\iota=0}^{k-j} \espp{\kappa_{\iota}}\times\espp{\norm{\tilde{y}}_{\EL_{p}[t_{j-1},t_j]}} \Big] .
\end{align*}

From the above inequality, using linearity of expectation and the discrete Young's inequality (see \cite[Lemma 1.1]{tanete07}) leads to
$\espp{\norm{W}_{\EL_p[0,t_{M}]}} \leq  \frac{s_M(\w)}{\w-L}\espp{\norm{\tilde{y}}_{\EL_p[0,t_M]}}$,
where $s_M(\w)\coloneqq  \sum_{j=0}^{M} \big(\frac{\w}{\w-L}\big)^j\times \prod_{\iota=0}^{j-1} \espp{\kappa_{\iota}}$.
To show $\EL_p$ stability in expectation of the error subsystem, it remains to prove that for $M\rightarrow\infty$ the series $s_\infty(\w)$ converges to a finite value. Since the underlying protocol is a.s. UGES, we have that $\espp{\kappa_{\iota}}\leq \bar{\kappa} < 1$ for all $\iota \in\N$. We thus get, for all $j\in\N$, $\big(\frac{\w}{\w-L}\big)^j\times \prod_{\iota=0}^{j-1} \espp{\kappa_{\iota}} \leq \big(\frac{\w\bar{\kappa}}{\w-L}\big)^j$.
Condition $(iii)$, i.e. $\w > L/(1-\bar{\kappa})$, ensures that the right-hand side is less than one. Therefore, $\sum_{j=0}^{\infty} \big(\frac{\w\bar{\kappa}}{\w-L}\big)^j = \frac{\w - L}{\w(1-\bar{\kappa})-L}$,
and by the comparison test for series \cite{knopp13}, this implies that $\sum_{j=0}^{\infty} \big(\frac{\w}{\w-L}\big)^j\times \prod_{\iota=0}^{j-1} \espp{\kappa_{\iota}}$
converges to a finite number which we denote by $s_{\infty}(\w)$. We have thus shown that $s_M(\w)$ converges to a finite number $s_{\infty}(\w)$ for $M\rightarrow\infty$. Therefore, 
\begin{align}\label{eq:Lp-proof-final}
	\espp{\norm{W}_{\EL_p[0,t_{M}]}} &\leq \frac{s_\infty(\w)}{\w-L}\espp{\norm{\tilde{y}}_{\EL_p[0,t_M]}}.
\end{align}
Either $\espp{\norm{\tilde{y}}_{\EL_p[0,t_M]}}=0$ or the ratio \sloppy $\espp{\norm{W}_{\EL_p[0,t_{M}]}}/\espp{\norm{\tilde{y}}_{\EL_p[0,t_M]}}$ is bounded by an expression that is independent of $M$, hence \eqref{eq:Lp-proof-final} remains true with $t$ in lieu of $t_M$ for any $t\geq 0$.
From Definition \ref{def:Lp-in-exp}, \eqref{eq:W0-contribution} and \eqref{eq:Lp-proof-final}, we conclude that the system \eqref{eq:e-sys}, \eqref{eq:protocol} is $\EL_p$ stable in expectation from $\tilde{y}$ to $W$ for $p\in\N\cup\{+\infty\}$ with finite gain $\gamma_e$.  \qedd

To conclude the proof of Theorem \ref{theo:Lp-overall}, Proposition \ref{propo:Lp-e-sys} is used together with the same small-gain arguments at the end of the proof of Theorem \ref{theo:stability-stoch-protocol}. This leads to the small gain condition \eqref{eq:omega-numeric} which determines the required transmission rate $\w$ that ensures stability. It only remains to show that this choice of $\w$ always exists. 
Note that $s_\infty(\w) \leq \frac{\w-L}{\w(1-\bar{\kappa})-L}$ since $\espp{\kappa_k}\leq \bar{\kappa}<1$. Therefore, $\w\to\infty$ is $\lim_{\w\to\infty} \frac{\gamma s_\infty(\w)}{\w - L} \leq \lim_{\w\to\infty} \frac{\gamma}{\w(1-\bar{\kappa})-L} = 0$ (since $\bar{\kappa}<1$). Moreover, $\gamma/(\w(1-\bar{\kappa})-L)$ is a continuous and  strictly decreasing function of $\w$ for $\w > L/(1-\bar{\kappa})$. Hence, $\exists \w^*>0$ such that for all $\w>\w^*$, $\gamma s_\infty(\w)/(\w-L)<1$, concluding the proof.\qedd

\section{Proof of Lemma \ref{lem:tabbara}}\label{sec:proof-Tabbara}
Given space limitations, we only show Lemma \ref{lem:tabbara} for the stochastic scheduling case since the deterministic scheduling case follows similarly.
The stability condition for stochastic scheduling is given by \eqref{eq:small-gain-condition}, and the counterpart in \cite{tabnes08-tac} is given by \eqref{eq:small-gain-condition} with $f_n=f$ for all $n\in\mathcal{N}$. Denote by $\text{SG}_{\text{\tiny Thm.\ref{theo:stability-stoch-protocol}}}$ the left-hand side of \eqref{eq:small-gain-condition}, and $\text{SG}_{\text{\tiny \cite{tabnes08-tac}}}$ the corresponding small-gain condition in \cite{tabnes08-tac}. We first show that $\text{SG}_{\text{\tiny Thm.\ref{theo:stability-stoch-protocol}}}\leq \text{SG}_{\text{\tiny \cite{tabnes08-tac}}}$ for $f=\min_{n\in\mathcal{N}}f_n$, which implies that \cite{tabnes08-tac} can ensure stability in our setting (i.e. $\text{SG}_{\text{\tiny \cite{tabnes08-tac}}}<1$ implies $\text{SG}_{\text{\tiny Thm.\ref{theo:stability-stoch-protocol}}}<1$), and then we show that both $\text{SG}_{\text{\tiny Thm.\ref{theo:stability-stoch-protocol}}}$ and $\text{SG}_{\text{\tiny \cite{tabnes08-tac}}}$ are strictly decreasing in $\w\in(0,\infty)$. These two conditions imply that $\w_{\text{\tiny Thm.\ref{theo:stability-stoch-protocol}}}\leq \w_{\text{\tiny \cite{tabnes08-tac}}}$. Note that $\sum_{n=1}^{N}\frac{N}{[N-(n-1)]f_n}\leq \sum_{n=1}^{N}\frac{N}{[N-(n-1)]f}$, where equality only holds in the case all probabilities are equal to the smallest one, i.e. $f_n=f,\forall n\in\mathcal{N}$. Moreover, we note that $\rho_{\w}$ in \eqref{eq:rho} is a strictly decreasing function in $f_n$, therefore, $\rho_{\w} \leq \rho_{\w}^{\text{\tiny \cite{tabnes08-tac}}}$ since $f\leq f_n$. Then, we have shown that $\text{SG}_{\text{\tiny Thm.\ref{theo:stability-stoch-protocol}}}\leq \text{SG}_{\text{\tiny \cite{tabnes08-tac}}}$ for $f=\min_{n\in\mathcal{N}}f_n$. That is, our small-gain condition is strictly smaller than the one in \cite{tabnes08-tac}, except for the case when $f_n=f,\forall n\in\mathcal{N}$, which is when our bounds are equal.
We next show that $\text{SG}_{\text{\tiny Thm.\ref{theo:stability-stoch-protocol}}}$ is strictly decreasing in $\w$.
Since $(1+\rho_{\w})/(1-\rho_{\w})$ is strictly increasing for $0<\rho_{\w}<1$, and $\rho_{\w}$ is strictly decreasing in $\w\in(0,\infty)$, we can conclude that $(1+\rho_{\w})/(1-\rho_{\w})$ is strictly decreasing in $\w\in(0,\infty)$, and so is $\text{SG}_{\text{\tiny Thm.\ref{theo:stability-stoch-protocol}}}$.\qedd

\section{Proof of Theorem \ref{theo:LTI-stoch}}\label{sec:proof-LTI-stoch} 
This proof boils down to proving that conditions $(ii)$, i.e. \eqref{eq:g-dynamics}, and $(iii)$ of Theorem \ref{theo:stability-stoch-protocol} hold with $A=\left|\overline{A}_{22}\right|$ and $\gamma = \sqrt{\bm{\mu}}$ if the LMI \eqref{eq:LMI-stoch} holds.
First, we show that an LTI system always satisfies \eqref{eq:g-dynamics}. That is, using $\Sigma_{\LTI}$ we can write for all $(x,e,w)\in\R^{n_x}\times\R^{n_e}\times\R^{n_w}$,
$\overline{\mathbf{g}}(x,e,w) = \overline{A_{21}x + A_{22}e + E_2w} \preceq  A\bar{e} + \tilde{y}(x,w)$,
where clearly $A=\left|\overline{A}_{22}\right|$ and $\tilde{y}(x,w)= G(x) + E\overline{w}$ with $G(x) = \overline{A_{21}x}$ and $E = \overline{E_2}$. 
Note that if there exists a Lyapunov function $V$ such that, for any $x\in\R^{n_x}$, $e\in\R^{n_e}$, and $w\in\R^{n_w}$,
\begin{multline}\label{eq:dVdt}
	\left\langle \nabla V(x), A_{11}x + A_{12}e+E_1w\right\rangle \leq -\varrho_1(|x|) - \varrho_2(|e|)\\ - |G(x)|^2 + \gamma^2 |e|^2 + \gamma^2 |w|^2
\end{multline}
holds with continuous and positive definite functions $\varrho_1,\varrho_2$, then the $x$-subsystem in $\Sigma_{\LTI}$ is $\EL_2$ stable from $(e,w)$ to $G(x)$ with finite gain $\gamma$. We then need to find $V$ such that \eqref{eq:dVdt} holds. The latter will lead to the LMI \eqref{eq:LMI} and the proof would be complete.
Let $V(x) \coloneqq  x^\top \textbf{P} x$ for any $x\in\R^{n_x}$. In view of $\Sigma_{\LTI}$, we have that $\left\langle \nabla V(x), A_{11}x + A_{12}e+E_1w\right\rangle = x^\top \left(\textbf{P}A_{11}+A_{11}^\top \textbf{P} \right) x  + x^\top \textbf{P}A_{12}e + x^\top \textbf{P}E_1w + e^\top A_{12}^\top \textbf{P}x + w^\top E_1^\top \textbf{P}x$.
On the other hand, by multiplying the right and left hand side of the LMI in \eqref{eq:LMI-stoch} by $(x,e,w)$ and its transpose, respectively, we obtain
$x^\top \left(\textbf{P}A_{11}+A_{11}^\top \textbf{P} \right) x  + x^\top \textbf{P}A_{12}e + x^\top \textbf{P}E_1w+ e^\top A_{12}^\top \textbf{P}x + w^\top E_1^\top \textbf{P}x \leq -\bm{\varepsilon}_1|x|^2 - \bm{\varepsilon}_2|e|^2 -|G(x)|^2  + \bm{\mu}|e|^2 + \bm{\mu}|w|^2$.
Consequently, \eqref{eq:dVdt} is verified with $\varrho_1(s)=\bm{\varepsilon}_1s^2$, $\varrho_2(s)=\bm{\varepsilon}_2s^2$, and $\gamma=\sqrt{\bm{\mu}}$. That is, if the LMI in \eqref{eq:LMI-stoch} holds, then the $x$-subsystem in $\Sigma_{\LTI}$ is $\EL_2$ stable from $(e,w)$ to $G(x)$. Since the plant is stabilisable and detectable, the feasibility of the LMI \eqref{eq:LMI-stoch} is ensured for sufficiently small $\bm{\varepsilon}_1,\bm{\varepsilon}_2$ and sufficiently large $\bm{\mu}$, see e.g. \cite[Remark 3]{abpoda16}.
The proof is thus complete.\qedd

\section{Proof of Proposition \ref{propo:equivalent-LP}}\label{sec:proof-equiv-LP}
The stability constraint \eqref{eq:stab-constraint} with $\Phi(\text{SINR}_i)=\exp(-a/\text{SINR}_i)$ can be written as
$\exp\big(-a \sum_{i=1}^{\ell} \frac{1}{\text{SINR}_i} \big) 
\geq \tfrac{\bar{\tau}(\gamma + L)}{1- \eta}  -\delta$.
We recall that $\text{SINR}_i>0$ for every $i\in\{1,\dots,\ell\}$. Since both sides of the above inequality are positive, we can apply natural logarithm and obtain $\sum_{i=1}^{\ell} \frac{1}{\text{SINR}_i} \leq \mathscr{C}_{\bar{\tau}}$. In terms of powers, this is equal to
\begin{align}\label{eq:stab-constraint-log}
	\frac{\sigma_1^2 + \sum_{j\neq i} g_{j1}p_j}{g_{11}p_1}+\cdots + \frac{\sigma_{\ell}^2 + \sum_{j\neq i} g_{j\ell}p_j}{g_{\ell\ell}p_\ell} \leq \mathscr{C}_{\bar{\tau}}.
\end{align}
Then, the proof of Proposition \ref{propo:equivalent-LP} boils down to showing that \eqref{eq:stab-constraint-log} is equivalent to
\begin{align}\label{eq:equiv-ineq}
	\smallmtx
	\begin{bmatrix}
		g_{11}q_1\mathscr{C}_{\bar{\tau}} & \cdots & -g_{\ell 1} \\
		\vdots & \ddots & \vdots \\
		-g_{1\ell} & \cdots & g_{\ell\ell}q_\ell \mathscr{C}_{\bar{\tau}} 
	\end{bmatrix}\begin{bmatrix}
		p_1\\ \vdots\\ p_\ell
	\end{bmatrix} \geq 
	\begin{bmatrix}
		\sigma_1^2 \\ \vdots \\ \sigma_\ell^2
	\end{bmatrix}
\end{align}
with $q\in\mathscr{Q}$. Obviously, \eqref{eq:equiv-ineq}$\Rightarrow$\eqref{eq:stab-constraint-log} by definition of $\mathscr{Q}$. Since each of the terms in \eqref{eq:stab-constraint-log} are strictly positive, we have that $\frac{\sigma_i^2+\sum_{j\neq i} g_{ji}p_j}{g_{ii}p_i} < \mathscr{C}_{\bar{\tau}}$ for every $i\in\{1,\dots,\ell\}$, and thus $\exists q\in\mathscr{Q}$ such that $\frac{\sigma_i^2+\sum_{j\neq i} g_{ji}p_j}{g_{ii}p_i} \leq q_i \mathscr{C}_{\bar{\tau}}$ for every $i\in\{1,\dots,\ell\}$, which leads to \eqref{eq:equiv-ineq}, concluding \eqref{eq:stab-constraint-log}$\Rightarrow$\eqref{eq:equiv-ineq}.
\qedd


\subsection{Proof of Proposition \ref{propo:optimisation}}\label{sec:proof-two-channel}
\begin{figure}[htb]
	\centering 
	\includegraphics[scale=0.9]{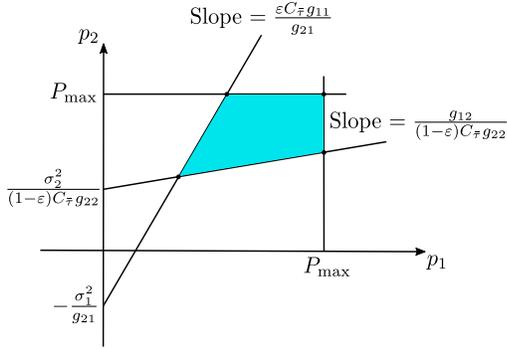}
	\caption{Polygon describing the constraint \eqref{eq:stab-constraint-2channel}.}
	\label{fig:polygon}
\end{figure}

For $\ell=2$, $q_1=\varepsilon$, and $q_2=1-\varepsilon$, \eqref{eq:LP-constraint} becomes
\begin{align}\label{eq:stab-constraint-2channel}
	\smallmtx
	\begin{bmatrix}
		g_{11}\varepsilon\mathscr{C}_{\bar{\tau}} & -g_{2 1} \\
		-g_{12} &  g_{22}(1-\varepsilon) \mathscr{C}_{\bar{\tau}} \\
		-1 &  0 \\
		0 &  -1
	\end{bmatrix}
	\mtxdu{p_1}{p_2}\geq 
	\begin{bmatrix}
		\sigma_1^2 \\  \sigma_2^2\\ -P_{\max} \\ -P_{\max}
	\end{bmatrix},
\end{align}
which determines the polygon in Fig. \ref{fig:polygon}. From the conditions of the Theorem we know that $\mathscr{C}_{\bar{\tau}}>0$. For the two-link LP, the optimal solution is given by the intersection of the two lines in Fig. \ref{fig:polygon}. However, we need them to intersect in the positive orthant, which happens if $\frac{\varepsilon \mathscr{C}_{\bar{\tau}}}{g_{21}}>\frac{g_{12}}{(1-\varepsilon)\mathscr{C}_{\bar{\tau}}g_{22}}$, or, equivalently, if $\varepsilon^2g_{11}g_{22}\mathscr{C}_{\bar{\tau}}^2-\varepsilon g_{11}g_{22}\mathscr{C}_{\bar{\tau}}^2 + g_{12}g_{21}<0$ (feasibility condition). Note that  $\mathscr{C}_{\bar{\tau}}>\sqrt{4g_{12}g_{21}/(g_{11}g_{22})}$ ensures the determinant of this polynomial is positive, and it translates to the upper bound on $\bar{\tau}$ in the proposition statement.
We can then factor this expression as $g_{11}g_{22}\mathscr{C}_{\bar{\tau}}^2(\varepsilon - \varepsilon_1)(\varepsilon - \varepsilon_2)<0$, where $	\varepsilon_1 = \frac{1}{2} + \sqrt{\frac{1}{4} - \frac{g_{12}g_{21}}{g_{11}g_{22}\mathscr{C}_{\bar{\tau}}^2}}, 
\varepsilon_2 = \frac{1}{2} - \sqrt{\frac{1}{4} - \frac{g_{12}g_{21}}{g_{11}g_{22}\mathscr{C}_{\bar{\tau}}^2}}$.
Therefore, the feasibility condition $\varepsilon^2g_{11}g_{22}\mathscr{C}_{\bar{\tau}}^2-\varepsilon g_{11}g_{22}\mathscr{C}_{\bar{\tau}}^2 + g_{12}g_{21}<0$ holds for $\varepsilon\in(\varepsilon_2,\varepsilon_1)$. The proof follows by finding the intersection of the lines $p_2=\frac{\varepsilon \mathscr{C}_{\bar{\tau}}}{g_{21}}p_1-\frac{\sigma_1^2}{g_{21}}$ and $p_2=\frac{g_{12}}{(1-\varepsilon)\mathscr{C}_{\bar{\tau}}g_{22}}p_1+\frac{\sigma_2^2}{(1-\varepsilon)\mathscr{C}_{\bar{\tau}}g_{22}}$, which leads to \eqref{eq:powers-epsilon}. The feasibility condition on the maximum power follows from ensuring the feasible set is non-empty, i.e. $P_{\max}\geq \max\{p_1^*(\varepsilon),p_2^*(\varepsilon)\}$, then the proof is complete. \qedd 

\bibliographystyle{abbrv}        
\bibliography{bibliography}

\end{document}